\documentclass[a4paper,11pt]{article}

\usepackage{longtable} 
\usepackage[square,sort,comma,numbers]{natbib}
\usepackage{etoolbox}
\usepackage[utf8x]{inputenc}
\usepackage{color,colortbl,xcolor}
\usepackage{bbding}
\usepackage{multirow}
\usepackage{lscape}
\usepackage{array,graphicx}
\usepackage{textcomp}
\usepackage{epstopdf}
\usepackage{threeparttable}
\usepackage{caption}
\usepackage{subcaption}
\usepackage{bbm}
\usepackage{nicefrac}
\usepackage{afterpage}
\usepackage{amsmath}
\newcommand{\ds}{}
\usepackage{setspace}
\usepackage{authblk}
\usepackage{amsfonts}
\usepackage{enumitem}
\usepackage{ulem}
\usepackage{booktabs}
\usepackage[colorinlistoftodos]{todonotes}

\usepackage{float}
\usepackage[margin=1in]{geometry}

\makeatletter
\renewcommand\paragraph{\@startsection{paragraph}{4}{\z@}%
	{-2.5ex\@plus -1ex \@minus -.25ex}%
	{1.25ex \@plus .25ex}%
	{\normalfont\normalsize\bfseries}}
\makeatother
\patchcmd{\thebibliography}{\chapter*}{\section*}{}{}

\definecolor{Gray}{gray}{0.95}

\definecolor{light-gray}{gray}{0.8}
\definecolor{very-light-gray}{gray}{0.9}

\newcommand{\Proxdelta}{Prox_{\mathbbm{1},\delta}}
\newcommand{\NB}[2]{\sout{#1}\textcolor{red}{#2}}

\pdfoutput=1

\begin{document}
\title{Metrics for describing dyadic movement: a review}

\author[1,2,*]{\small Rocio Joo}
\author[3]{\small Marie-Pierre Etienne}
\author[4]{\small Nicolas Bez}
\author[2]{\small St\'ephanie Mah\'evas}

\affil[1]{\footnotesize Department of Wildlife Ecology and Conservation, Fort Lauderdale Research and Education Center, University of Florida, Fort Lauderdale, FL, USA}
\affil[2]{\footnotesize IFREMER, Ecologie et Mod\`eles pour l'Halieutique, BP 21105, 44311 Nantes Cedex 03, France}
\affil[3]{\footnotesize Univ Rennes, Agrocampus Ouest, CNRS, IRMAR - UMR 6625, F-35000 Rennes, France}
\affil[4]{\footnotesize MARBEC, IRD, Ifremer, CNRS, Univ Montpellier, S\`ete, France}
\affil[*]{Corresponding author: Rocio Joo, rocio.joo@ufl.edu}

\date{}

\maketitle 

Running headline: Review of metrics for joint movement 

\newpage

\begin{abstract}
 In movement ecology, the few works that have taken collective behaviour into account are data-driven and rely on simplistic theoretical assumptions, relying in metrics that may or may not be measuring what is intended. In the present paper, we focus on pairwise joint-movement behaviour, where individuals move together during at least a segment of their path. We investigate the adequacy of twelve metrics introduced in previous works for assessing joint movement by analysing their theoretical properties and confronting them with contrasting case scenarios. Three criteria are taken into account for review of those metrics: 1) practical use, 2) dependence on parameters and underlying assumptions, and 3) computational cost. When analysing the similarities between the metrics as defined, we show how some of them can be expressed using general mathematical forms. In addition, we evaluate the ability of each metric to assess specific aspects of joint-movement behaviour: proximity (closeness in space-time) and coordination (synchrony) in direction and speed. We found that some metrics are better suited to assess proximity and others are more sensitive to coordination. To help readers choose metrics, we elaborate a graphical representation of the metrics in the coordination and proximity space based on our results, and give a few examples of proximity and coordination focus in different movement studies.  
\end{abstract}


 
\section{Introduction}

Collective behaviour has been the object of study of many disciplines, such as behavioural ecology, psychology, sports, medicine, physics and computer sciences \cite{Sumpter2012,Biro2016,Duranton2016,Conradt2009,Travassos2013}. In multiple contexts, individuals -- in a very wide sense of the word -- adapt their behaviour as a function of their interaction with others. In movement ecology, where movement is regarded as an expression of behaviour \cite{Nathan2008}, collective behaviour should be considered as a key element given that collective dynamics and individual movement are intricately intertwined \cite{Biro2016}. Accordingly, mechanistic movement models should account for these dynamics. The vast majority of movement models neglect this aspect, with a few exceptions \cite[e.g.][]{Langrock2014,Potts2014,Niu2016, Russell2016}. The consequence has been that the forms that these dynamics take in the few existing works rely on very simple theoretical assumptions.

Collective behaviour can be produced at large group scales (flocks, colonies, schools) but also at small group scales (triads, dyads). Regardless of the actual group scale, global patterns of collective behaviour originate from local interactions among neighbouring members \cite{Camazine2003}, so analysing dyad interaction as a first step is a pertinent choice. Concerning dyadic interaction, here we focus on what we call `joint movement', where two individuals  move together during the total duration or a partial segment of their paths. Dyadic movement behaviour has been mostly studied in a data-driven approach, using several metrics to quantify it. In movement ecology, few works have applied and compared some of these metrics \cite{Miller2012,Long2014}. However, their theoretical properties, and thus the similarities and differences in their construction and in what they actually assess, have not been thoroughly analysed yet. 

This manuscript reviews a series of metrics used to assess pairwise joint-movement and proposes some modifications when appropriate (Table \ref{IndReview}). Three criteria are taken into account for the review of these metrics: practical use, dependence on parameters and computational cost; they are evaluated through both a theoretical (conceptual) as well as a practical approach. Metrics found in the literature essentially measured two aspects of joint movement: proximity and coordination. Proximity refers to closeness in space-time, as in how spatially close simultaneous fixes are in a dyad (a point pattern perspective). The notion of proximity is thus subjective, since a judgement on proximity involves a threshold in distance whether local or global, or the definition of a reference zone (where encounters may be observed). Coordination, on the other hand, refers to synchrony in movement, which can be assessed through measures of similarity or correlation in movement patterns such as speed or direction. There might be a thin line between proximity and coordination, and some metrics may be associated with both at some degree, as we show through the description of their theoretical properties and the practical analysis of case scenarios. 

The manuscript is thus organized as follows. We first describe the criteria used to evaluate the metrics as indices of dyadic joint movement. We then present the different metrics and their theoretical properties with special attention to their dependence towards parameters. Next, we define case scenarios to evaluate the practical properties of the metrics. After that, we introduce a small section of simple dyad simulation to evaluate the computational cost of the metrics. In the last section, we discuss the overall suitability of the metrics for assessing joint movement in ecology and give some practical guidelines for their use. 

\section{Evaluation criteria}
\label{Evaluation criteria}

We categorized the desirable properties of metrics for assessing dyadic joint movement into three criteria: practical use, considered the most important one; dependence on parameters; and computational cost: 

\begin{enumerate}[label=C\arabic*]
	\item Practical use \cite{Rice2005,Rochet2003,VanStrien2012}: 1) A metric is useful if it is interpretable and reflects a marked property of collective behaviour. 2) It should also be sensitive to changes in patterns of joint movement (e.g. higher values for high joint movement and lower values for independence in movement). 3) Being able to attain the theoretical range of values would also be important, as not doing so makes it harder to interpret empirical values. C1 is therefore a three dimensional criterion comprising interpretation, sensitivity and attainable range. Attainable range is covered in the theoretical properties section; we highlight the difficulties or implausibility to attain minimum and maximum values for the metrics when this is true. How to interpret each metric is also explained in this section; evidently, a metric without an attainable range is difficult to interpret. Sensitivity is addressed in the case-scenario section. \label{C:Use}
	\item Dependence on parameters: A metric that depends on few parameters and hypotheses is more robust and generic than one that strongly relies on many parameters and hypotheses, since the former can produce more easily comparable results and interpretations. In addition, an ideal metric can be defined in such a way that the user can easily see how a change in the values of the parameters or in the components related to movement assumptions conditions the metric derivations and interpretations. In the next section, we describe the assumptions underlying each metric and the parameters needed to be fixed by the user. This description will allow distinguishing user-tractable parameter-dependent metrics from those that are not. \label{C:Param}
	\item Computational cost: It may be the least important criterion, but ideally a metric should not take much computational time to be derived, especially when processing numerous large trajectories. This criterion is evaluated in section \ref{CPUtime} through the simulation of dyads and calculation of the computational time required to derive each metric. \label{C:Cost}
\end{enumerate}


\section{Definition and theoretical properties of the metrics} \label{Indices}

In the following subsections the metrics are defined and their theoretical properties are described.  A summary is proposed in Table \ref{IndReview}. Considering two individuals named $A$ and $B$, the position of $A$ (resp. $B$) at time $t$ is denoted by $X_t^A$ (resp. $X_t^B$). The distance between $A$ at time $t_1$ and $B$ at time $t_2$ will be referred to as $d_{t_1,t_2}^{A,B}$. When the distance between two individuals is regarded at simultaneous time, this will be shortened to $d_t^{A,B}$. Whenever possible, metrics introduced by different authors but that are actually very similar in their definition, are grouped \NB{here}{} under a unified name and a general definition. 

	\begin{table}[ht!]
		\caption{Metrics for measuring dyad joint movement}
		\begin{center}
			\begin{tabular}{>{\centering\arraybackslash}p{10.3cm}p{1.7cm}p{4cm}}
				\hline
				\hline
				\rule{0pt}{11pt}
				{Metric} & {Range} & Parameters fixed ad hoc and Assumptions \\
				\hline 
				& &   \\				
				{$Prox = K_{\delta}^+/T$} & [0 , 1] & i) $\delta$: distance threshold, ii) K : kernel  \\
				{$Cs = \ds\frac{ D_{chance} - \Big(\ds\sum_{t=1}^{T} d_{t}^{A,B} \Big)/T
				}{D_{chance} + \Big(\ds\sum_{t=1}^{T} d_{t}^{A,B} \Big)/T} $} & $]-1 , 1]$ & $D_{chance}$ definition \\
				\multirow{2}{*}{$HAI = \ds\frac{K_{\delta}^+}{K_{\delta}^+ + (n_{A0} + n_{0B})/2}$} & \multirow{2}{*}{$[0  , 1]$} & i) Reference area, ii) $\delta$: distance threshold \\
				{$L_{ixn}T = \text{logistic}\left(\ln{\bigg( \ds\frac{n_{AB}/p_{AB} + n_{00}/p_{00}}{n_{A0}/p_{A0} + n_{0B}/p_{0B} }\bigg)}\right)$} & [0 , 1] & Reference area;  \\				                
                \multirow{2}{*}{$jPPA = \ds\frac{S\left\{ \bigcup\limits_{t=1}^{T-1} \left( E_{\phi^A}(X_t^A,X_{t+1}^A) \cap E_{\phi^B}(X_t^B,X_{t+1}^B) \right) \right\}}{S\left\{ \bigcup\limits_{t=1}^{T-1} \left( E_{\phi^A}(X_t^A,X_{t+1}^A) \cup E_{\phi^B}(X_t^B,X_{t+1}^B) \right) \right\}}$} & \multirow{2}{*}{[0 , 1]} &  i) Every zone within  ellipse has same odd  of being transited, ii) $\phi$: maximum velocity \\
                & & \\
                \multirow{2}{*}{$CSEM = \frac{\max \left\{m; N_m  >0\right\} }{T-1}$} & \multirow{2}{*}{[0 , 1]} & distance \\
                & & threshold \\
				\multirow{2}{*}{ $r_V = \frac{\ds\sum_{t=1}^{T}(V^A_t - \bar{V}^A)(V^B_t - \bar{V}^B)}{\sqrt{\ds\sum_{t=1}^{T}(V^A_t - \bar{V}^A)^2}\sqrt{\ds\sum_{t=1}^{T}(V^B_t - \bar{V}^B)^2}}$}
					 & \multirow{2}{*}{[-1 , 1]} &   \\
				& &  \\
				\rule{0pt}{11pt}
				\multirow{2}{*}{$DI_d = \left(\ds\sum_{t=1}^{T-1} \left[1 - \left(\frac{\mid d_{t,t+1}^{A}-d_{t,t+1}^{B}\mid}{d_{t,t+1}^{A}+d_{t,t+1}^{B}}\right)^\beta\right]\right)/(T-1)$} & \multirow{2}{*}{[0 , 1]} & $\beta$: scaling \\		
				& & parameter \\
                & &  \\
				\multirow{2}{*}{$DI_\theta = \left(\ds\sum_{t=1}^{T-1} \cos(\theta_{t,t+1}^{A} - \theta_{t,t+1}^{B})\right)/(T-1)$} & \multirow{2}{*}{[-1 , 1]} &  \\		
				& &  \\
				\multirow{4}{*}{$DI = \ds\frac{\ds\sum_{t=1}^{T-1} \cos(\theta_{t,t+1}^{A} - \theta_{t,t+1}^{B})\left[1 - \left(\frac{\mid d_{t,t+1}^{A}-d_{t,t+1}^{B}\mid}{d_{t,t+1}^{A}+d_{t,t+1}^{B}}\right)^\beta\right]}{T-1}$} & \multirow{4}{*}{[-1 , 1]} & \\
				& & $\beta$: scaling \\		
				& & parameter \\
				& &  \\
				\hline
				\hline
			\end{tabular}
		\end{center}
		\begin{tablenotes}
			\small
			\item \textit{Note:} The formulas assume simultaneous fixes. $K_{\delta}^+ = \ds\sum_{t=1}^T K_{\delta}(X^{A}_{t},X^{B}_{t})$; T is the number of (paired) fixes in the dyad; $\delta$ is a distance-related parameter. K is a kernel function. $A$, $B$: the two individuals in the dyad; $T$: number of fixes in the dyad; $D_{chance}$ is the chance-expected distance between A and B; $n_{AB}$: number of observed fixes where $A$ and $B$ are simultaneously in the reference area (when a subscript is $0$, it represents the absence of the corresponding individual from the reference area); $p_{AB}$: probability of finding $A$ and $B$ simultaneously in the reference area (same interpretation as for $n$ when a subscript is $0$); $E_{\phi^A}(X_t^A,X_{t+1}^A)$ is the ellipse formed with positions $X_t$ and $X_{t+1}$, and maximum velocity $\phi$ from individual $A$ (analogous for B); $S{}$ represents the surface of the spatial object between braces; $V^A$ (and $V^B$, resp.) represents the analysed motion variable of A (and B); $\bar{V}^A$ (and $\bar{V}^B$) represent their average; $\beta$ is a scale parameter; $\theta$, the absolute angle; $N_m$ is the number of m-similar consecutive segments within the series of analysed steps. 
		\end{tablenotes}
		\label{IndReview}
	\end{table}

\subsection{Proximity index (Prox)}
\label{subsec:prox}
\begin{figure}[ht!]
\centering
		\includegraphics[scale=0.35]{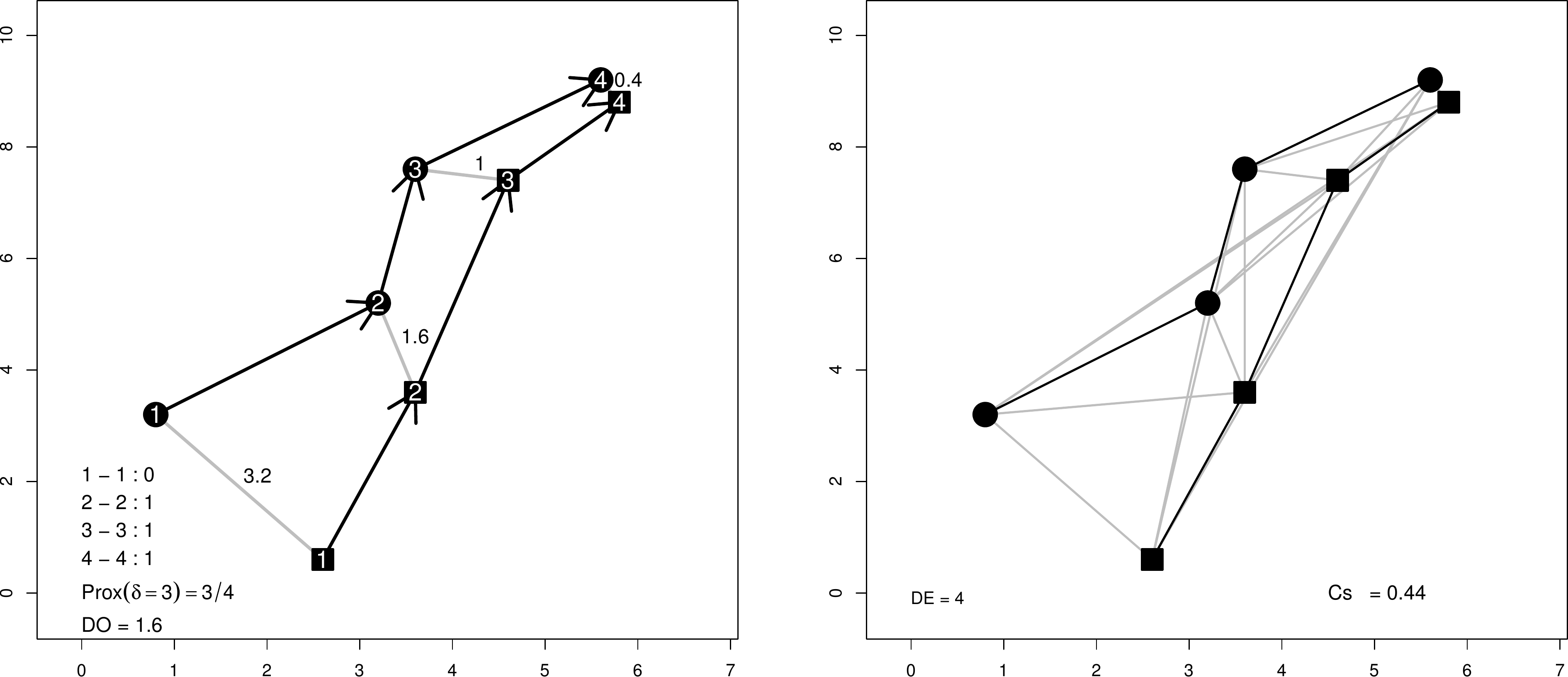}
	\caption{Example of Prox for $\delta=3$ (left panel) and Cs  (right panel). Circles and squares represent locations of two different individuals. Left panel: The numbers inside as well as the arrows represent the time sequence of both tracks. Grey lines correspond to the distances between simultaneous fixes; their values are shown. At the bottom: a dummy variable indicating if distances are below $\delta$ for each pair of simultaneous fixes, then the derived Prox and $D_O$ (average of observed distances). Right panel: Grey lines represent the distances of all permuted fixes; $D_E$ is their average.}
	\label{ProxCsEx}
\end{figure}

The proximity index (Prox in \cite{Bertrand1996}) is defined as the proportion of simultaneous pairs of fixes within a distance below an ad hoc threshold (Fig. \ref{ProxCsEx}). Other metrics in the literature are actually analogous to Prox: the coefficient of association (Ca) \cite{Cole1949} and the $I_{AB}$ index \cite{Benhamou2014}. Denoting by $T$ the number of pairs of fixes in the dyad, we propose a unified version of those metrics using a kernel $K$ (formula \ref{ProxEq1}):

\begin{equation}
Prox_{K,\delta} = \frac{1}{T} \sum_{t=1}
^T K_{\delta}(X_t^A, X_t^B), 
\label{ProxEq1}
\end{equation}

where $\delta$ is a distance threshold parameter.

Choosing $K_{\delta}(x,y) =\mathbbm{1}_{\{ \| x-y \| < \delta \}}$  ($\mathbbm{1}_{ \{\}}$ represents the indicator function)  as a kernel leads to the Prox metric in \cite{Bertrand1996}, denoted by $\Proxdelta$ henceforward. Instead, choosing $K_{\delta}(x,y)=\exp\big(- \| x-y \|^2  /(2 \delta^2)\big)$ gives the $I_{AB}$ index. Regarding Ca, for simultaneous fixes, its definition becomes exactly the same as $\Proxdelta$ (using Ca's adaptation to wildlife telemetry data shown in \cite{Long2014}).

Most of the proximity-related metrics are based on symmetric kernels and depend only on the distance between $A$ and $B$; therefore, the formula notation (\ref{ProxEq1}) can be simplified as:
\begin{equation}
Prox_{K,\delta} = \frac{1}{T} \sum_{t=1}
^T K_{\delta}(X_t^A, X_t^B) = \frac{1}{T}  K_{\delta}^+.
\label{ProxEq2}
\end{equation}

If the distance between two individuals is below the threshold $\delta$ during their whole tracks, $\Proxdelta$  will be 1 (and 0 in the opposite case). $\Proxdelta$ might be interpreted as the proportion of time the two individuals spent together. This interpretation is, of course, threshold dependent. The $I_{AB}$ index provides a smoother measure of the average proximity between two individuals along the trajectory. Proximity is thus dependent on the choice of a $\delta$ parameter and of a kernel function. 
Graphical examples illustrating the differences in $K_{\delta}(x,y) =\mathbbm{1}_{\{ \| x-y \| < \delta \}}$ and $K_{\delta}(x,y)=\exp\big(- \| x-y \|^2  /(2 \delta^2)\big)$ are in appendix \ref{appendix:ProxDif}. 

\subsection{Coefficient of Sociality (Cs) }
 The Coefficient of Sociality (Cs) \cite{Kenward1993} compares the mean distance between simultaneous pairs of fixes ($D_O$) against the mean distance between all permutations of all fixes ($D_E$).
 
\begin{equation}
Cs = \ds\frac{ D_E - D_O
}{D_E + D_O} = 1 - 2 \frac{  D_O
}{D_E + D_O} ,
\label{CsEq}
\end{equation}
where 
\begin{equation*}
D_O = \Big(\ds\sum_{t=1}^{T} d_t^{A,B} \Big)/T,
\end{equation*}
and
\begin{equation*}
D_E = \Big(\ds\sum_{t_1=1}^{T}\ds\sum_{t_2=1}^{T}d_{t_1,t_2}^{A,B}\Big)/T^2.
\end{equation*}

\cite{Kenward1993} stated that $Cs$ belongs to $[-1,1]$, and it has been used as a symmetrical index since. Nevertheless, that is not true. $Cs$ equals $1$  if and only if  $D_O=0$ and $D_E\ne0$, which occurs only when the two individuals always share the exact same locations. However, $Cs$ equals $-1$, if and only if $D_E=0$ and $D_O \neq 0$, which is impossible. $Cs$ equals $0$ when $D_O=D_E$. 

If all simultaneous fixes are very proximal but not in the same locations, $Cs$ would approach $1$ (how close to $1$ would depend on the value of $D_E$ as illustrated in the right hand side of equation \ref{CsEq}).  Moreover, only if $D_E < D_O$, $Cs$ can take a negative value. For $Cs$ to take a largely negative value, the difference in the numerator should be very large regarding the sum in the denominator; in appendix \ref{appendix:Cs1Negative} we show how implausible that situation is and how sensitive it is to the length of the series. The latter makes Cs from dyads of different length difficult to compare, because their real range of definition would differ.
This fact is neither evoked in the work that introduced the metric \cite{Kenward1993} nor in the ones that evaluated this and other metrics \cite{Miller2012,Long2014}, despite the fact that in those works no value lower than $-0.1$ was obtained.

Indeed, \cite{Kenward1993} assumed that the permutation of all fixes is a way to represent locations of independent individuals.
While this is questionable, some modified versions, as the one proposed by \cite{White1994}, use correlated random walks as null models and simulated independent trajectories under these models to replace $D_E$ by a more realistic  reference value.
Thus, a generalized version of $Cs$ would be:
\begin{equation}
Cs = \ds\frac{ D_{chance} - D_O}{D_{chance} + D_O},
\label{CsEqGen}
\end{equation}
where $D_{chance}$ is  defined through a user-chosen movement model for independent trajectories.


\subsection{The Half-weight Association Index (HAI)}
\label{sectionHAI}

The Half-weight Association Index (HAI) proposed by \cite{Brotherton1997} measures the proportions of fixes where individuals are close to each other (within a user-defined threshold). By that definition, HAI is exactly the same as $\Proxdelta$. However, HAI was popularized by \cite{Atwood2003} in another form that did not consider all fixes for the computation of the metric, but used counts with respect to a reference area (called overlapping zone in the original paper):

\begin{equation}
HAI = \frac{K_{\delta}^+}{K_{\delta}^+ + \frac{1}{2}(n_{A0}+n_{0B})}
\label{HAIEq}
\end{equation}
where $n_{AB}$ (resp $n_{A0}$; $n_{0B}$; $n_{00}$) is the number of simultaneous occurrences  of A and B in the reference area $S_{AB}$ (resp. simultaneous presence of A and absence of B; simultaneous absence of A and presence of B; simultaneous absence of A and absence of B), and where $K_{\delta}^+$ is computed over the reference area. 

It is worth noticing that the HAI adaptation proposed by \cite{Atwood2003} does not correctly account for spatial joint movement, as would do a $\Proxdelta$ version constraint to the reference area; i.e. the denominator should be equal to $n_{AB}$ + $n_{A0}$ + $n_{0B}$, which is the total number of simultaneous fixes where at least one individual is in the reference area. 

The dependence to the definition of an overlapping zone or reference area is discussed in the following subsection dedicated to $L_{ixn}T$, which also relies on the definition of a static reference area.

If the individuals remain together (i.e. in the reference area and closer than $\delta$) all the time, HAI is close to 1, and 0 in the opposite case. An example of the computation of HAI under \cite{Atwood2003}'s definition is given in Fig. \ref{LixnHAIEx}.

\subsection{Coefficient of Interaction ($L_{ixn}$ and $L_{ixn}T$)} 

\cite{Minta1992} proposed a Coefficient of Interaction ($L_{ixn}$) that assesses how simultaneous are the use and avoidance of a reference area $S_{AB}$ by two individuals:

\begin{equation}
L_{ixn} = \ln{\bigg( \frac{n_{AB}/p_{AB} + n_{00}/p_{00}}{n_{A0}/p_{A0} + n_{0B}/p_{0B} }\bigg)},
\label{LixnEq}
 \end{equation}
 
 where $p_{AB}$ is the probability, under some reference null model, of finding $A$ and $B$ simultaneously in $S_{AB}$ (the same interpretation as for $n$ when a subscript is 0; see subsection \ref{sectionHAI}). Attraction between individuals would cause greater simultaneous use of $S_{AB}$ than its solitary use, which would give positive values of $L_{ixn}$. Conversely, avoidance would translate into negative values of $L_{ixn}$, since use of $S_{AB}$ would be mostly solitary. A logistic transformation of the metric ({$L_{ixn}T$}) produces values between $0$ (avoidance) and $1$ (attraction), making the interpretation easier:
{\begin{equation}
L_{ixn}T = logistic(L_{ixn})= \frac{1}{1+e^{-L_{ixn}}}.
\label{LixnTEq}
\end{equation}}

Minta  proposed two different approaches for computing the associated probabilities conditionally to the fact that the reference area  is known (see examples in Fig. \ref{LixnHAIEx} and Table in appendix \ref{appendix:pLixn}). In both cases, the probabilities are estimated under the assumptions of independence in movement among the individuals and of uniform utilization of the space. Indeed this latter assumption can be relaxed and $p_{AB}$ can be derived from any kind of utilization distribution (see for instance \cite{Fleming2016} for the estimation of utilization distribution). 

\begin{figure}[ht!]
	\centering%
	\includegraphics[scale=0.35]{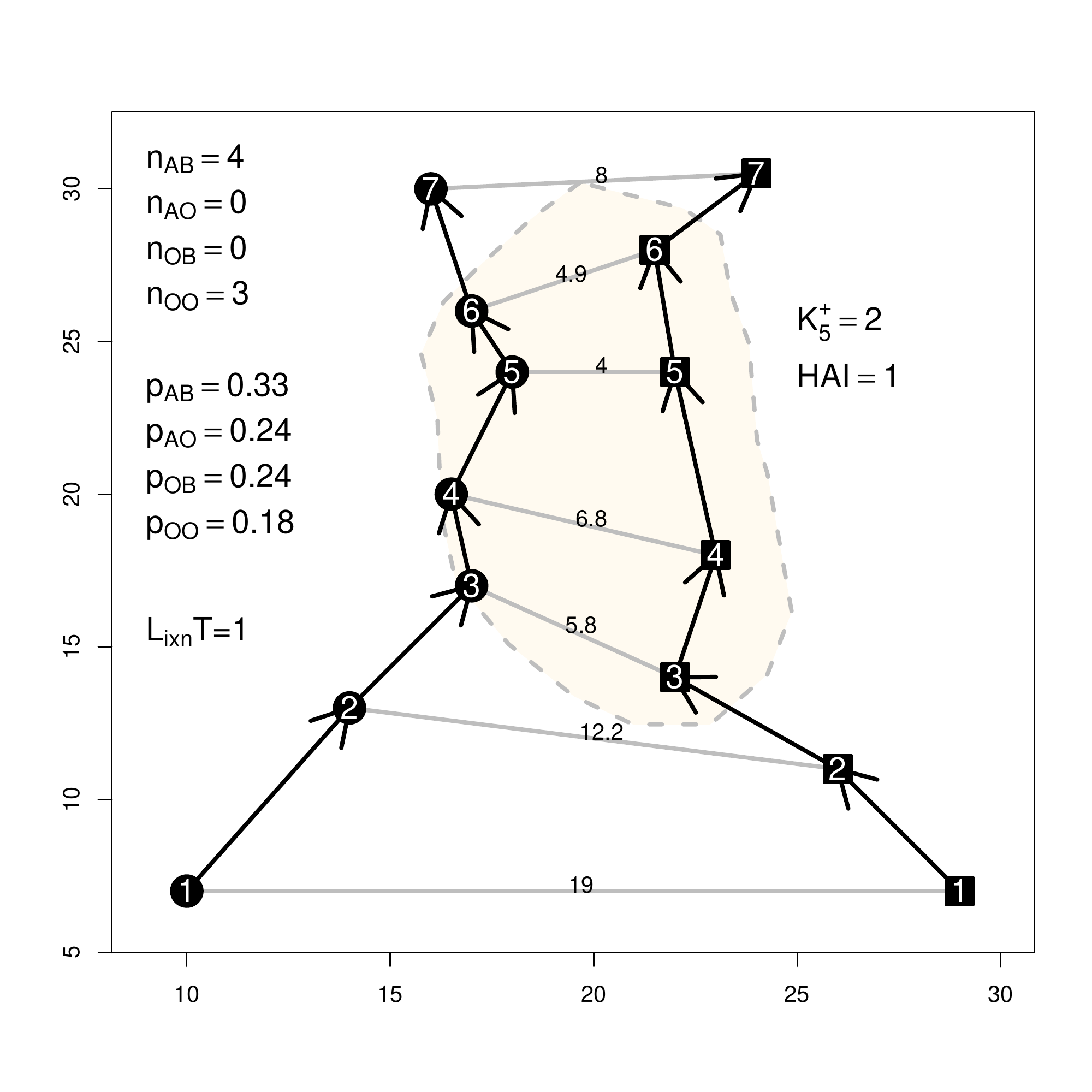}
    \includegraphics[scale=0.35]{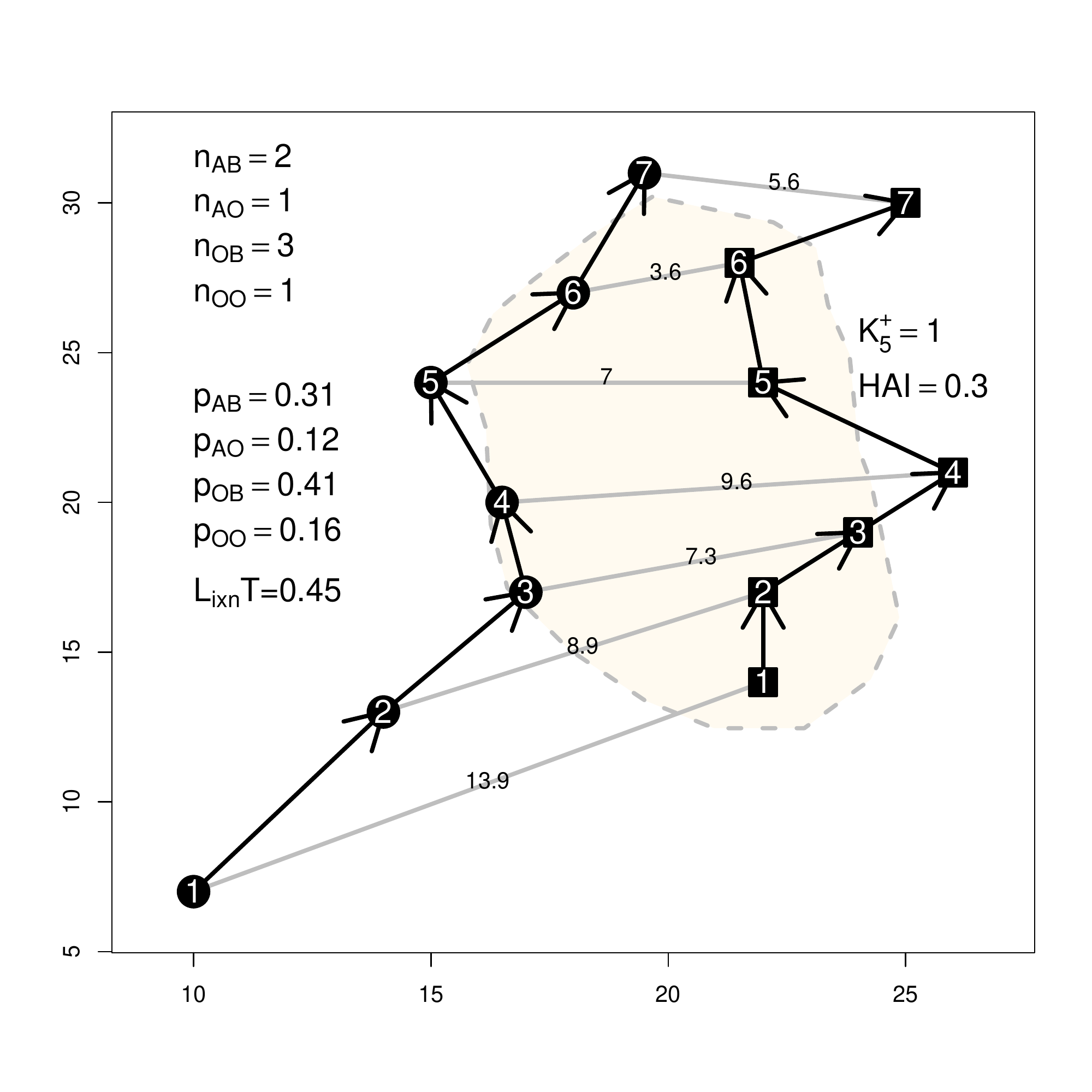}
	\caption{Two examples of the derivation of $L_{ixn}T$ and HAI. $L_{ixn}T$ was computed using expected frequencies. HAI was computed with $K_{\delta}(t) = \mathbbm{1}{\{ d_t^{A,B} < 5 \}}$. Circles and squares represent locations of two different individuals (A and B). The numbers inside as well as the arrows represent the time sequence of both tracks. Grey lines correspond to the distances between simultaneous fixes; their values are shown. The dashed lines circle an arbitrary reference area. }
	\label{LixnHAIEx}
\end{figure}

HAI and $L_{ixn}T$ rely heavily on a static reference area -- either known or estimated -- and on the probabilities of presence within this {reference} area. The static {reference} area could be defined, for instance, as the intersection of the respective home ranges of A and B. However, there are many approaches for estimating home ranges, each one relying on particular assumptions about the spatial behaviour of the studied populations \cite{Borger2008}. Thus, $S_{AB}$ is not a simple tuning parameter. The way it is defined may completely modify the output. If the reference area is equal to the whole area of movement of the two individuals, then both the numerator and the denominator in the logarithm are equal to infinity and $L_{ixn}T$ cannot be derived. That problem could arise for extremely mobile individuals, such as tuna, turtles and seabirds \cite{Block2011}, or fishing vessels \cite{Bertrand2008}, and avoiding it would require the computation of multiple dynamic reference areas. Therefore, $L_{ixn}T$ may be better used for specific cases where the definition of the {reference} area relies on a deep knowledge of the spatial behaviour of the populations.

\subsection{Joint Potential Path Area (jPPA)} 
\begin{figure}[ht!]
	\centering%
	\includegraphics[scale=0.4]{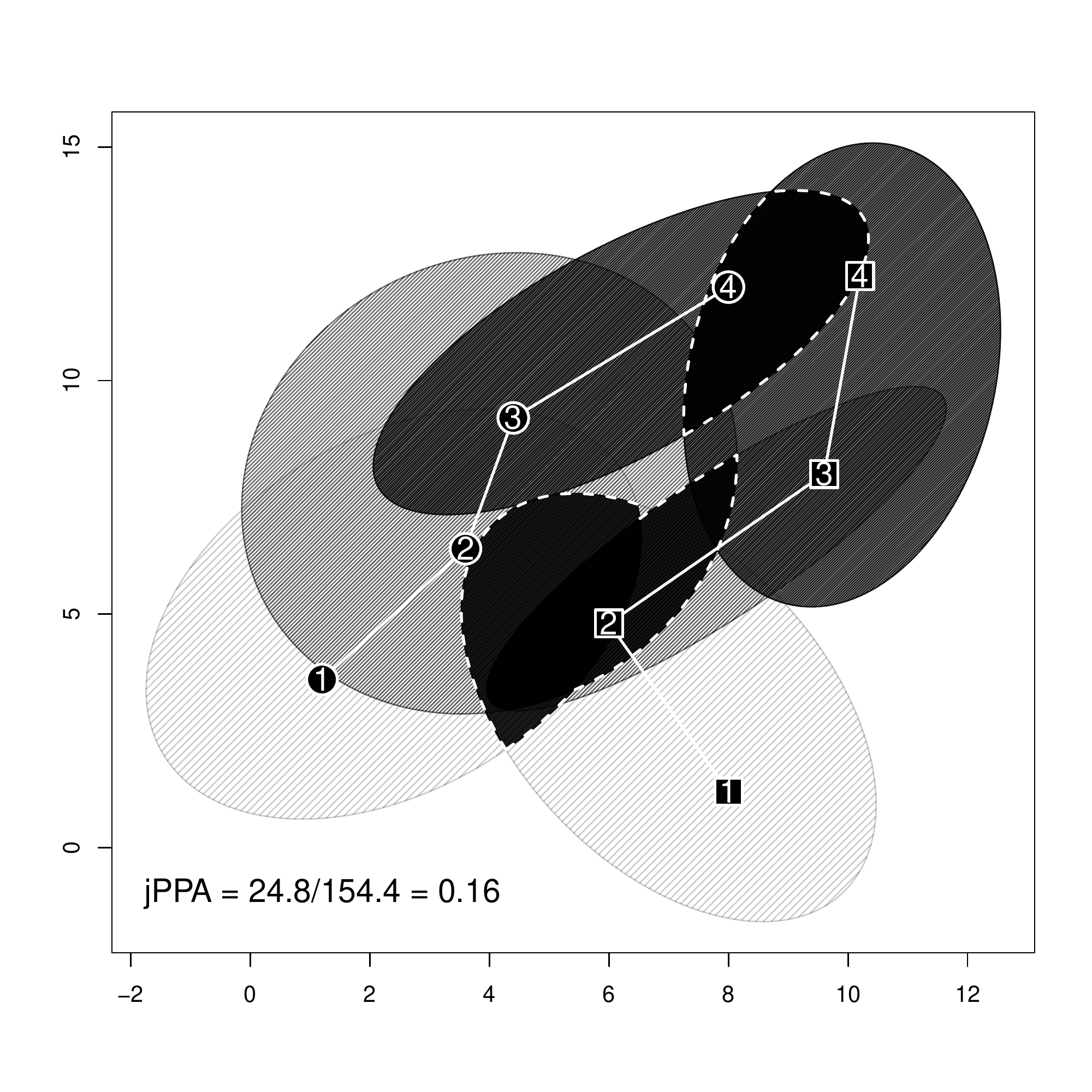}
	\caption{Example of the derivation of the joint potential path area (when $\phi=10$). Circles and squares represent locations of two different individuals (A and B); the numbers inside represent the time sequence. The grey scales of the ellipses correspond to the time intervals used for their computation: from light grey for the $[1,2]$ interval to dark grey for the $[3,4]$ interval. The black regions with white dashed borders correspond to the potential meeting areas.}
	\label{jPPAEx}
\end{figure}

\cite{Long2015a} computed the relative size of the potential encounter area at each time step of two individuals' tracks. 
Assuming a speed limit $\phi$, the potential locations visited between two consecutive fixes define an ellipse (appendix \ref{appendix::Ellipse}). Then, the potential encounter area corresponds to the intersection between the ellipses of the two individuals (at simultaneous time steps; see Fig. \ref{jPPAEx}). The overall potential meeting area is given by the spatial union of all those potential encounter areas. This area is then normalized by the surface of  the spatial union of all the computed ellipses to produce the joint Potential Path Area (jPPA) metric ranging from 0 to 1 (see formula in Table \ref{IndReview}). 
jPPA values close to 0 indicate no potential spatio-temporal overlap, while values close to 1 indicate a strong spatio-temporal match. 

Several issues can be discussed here. 
First, no movement model is assumed and therefore the method confers the same probabilities of presence to every subspace within the ellipse regions. This is clearly unrealistic as individuals are more likely to occupy the central part of the ellipse because they cannot always move at $\phi$, i.e. maximal speed. Second, the computation of the ellipses relies strongly on the $\phi$ parameter. 
If $\phi$ is unrealistically small, it would be impossible to obtain the observed displacements and the ellipses could not be computed. By contrast, if $\phi$ is too large, the ellipses would occupy such a large area that the intersected areas would also be very large (hence a large jPPA value). Alternatively, \cite{Long2015d} proposed a dynamic computation of $\phi$ as a function of the activity performed by the individual at each fix. 
Within this approach, additional information or knowledge (i.e. other data sources or models) would be required for the computation of $\phi$.

%
\subsection{Cross sampled entropy (CSE and CSEM)} Cross sampled entropy (CSE) \cite{Richman2000} comes from the time series analysis literature and is used for comparing pairs of motion variables \cite[e.g.][]{Duarte2013,Barnabe2016}. It evaluates the similarity between the dynamical changes registered in two series of any given movement measure. Here we present a simplification of the CSE for simultaneous fixes and position series. A segment of track A would be said to be $m-$similar to a segment of track B  if the distance between paired fixes from A and B remain below a certain threshold during  $m$ consecutive time steps. If we define $N_m$ as the number of $m-$similar segments within the series, then CSE can be defined as (the negative natural logarithm of) the ratio of $N_{m+1}$ over $N_m$ and might be understood as (the negative natural logarithm of) the probability for an $m-$similar segment to also be $(m+1)$-similar. Formally, CSE is defined as:

\begin{equation}
CSE_{\delta}(m) = -\ln\left\{\ds\frac{ \ds\sum_{t=1}^{T-m} \mathbbm{1}{\{ (\max_{k \in [0,m]} \mid X_{t+k}^{A} - X_{t+k}^{B} \mid ) < \delta \} }
}{\ds\sum_{t=1}^{T-m} \mathbbm{1}{\{ (\max_{k \in [0,m-1]} \mid X_{t+k}^{A} - X_{t+k}^{B} \mid ) < \delta \} }}\right\}=-\ln\frac{N_{m+1}}{N_m}
,
\label{CrossEq2}
\end{equation}

A large value of CSE corresponds to greater asynchrony between the two series, while a small value corresponds to greater synchrony.

\begin{figure}[ht!]
	\centering%
	\includegraphics[scale=0.4]{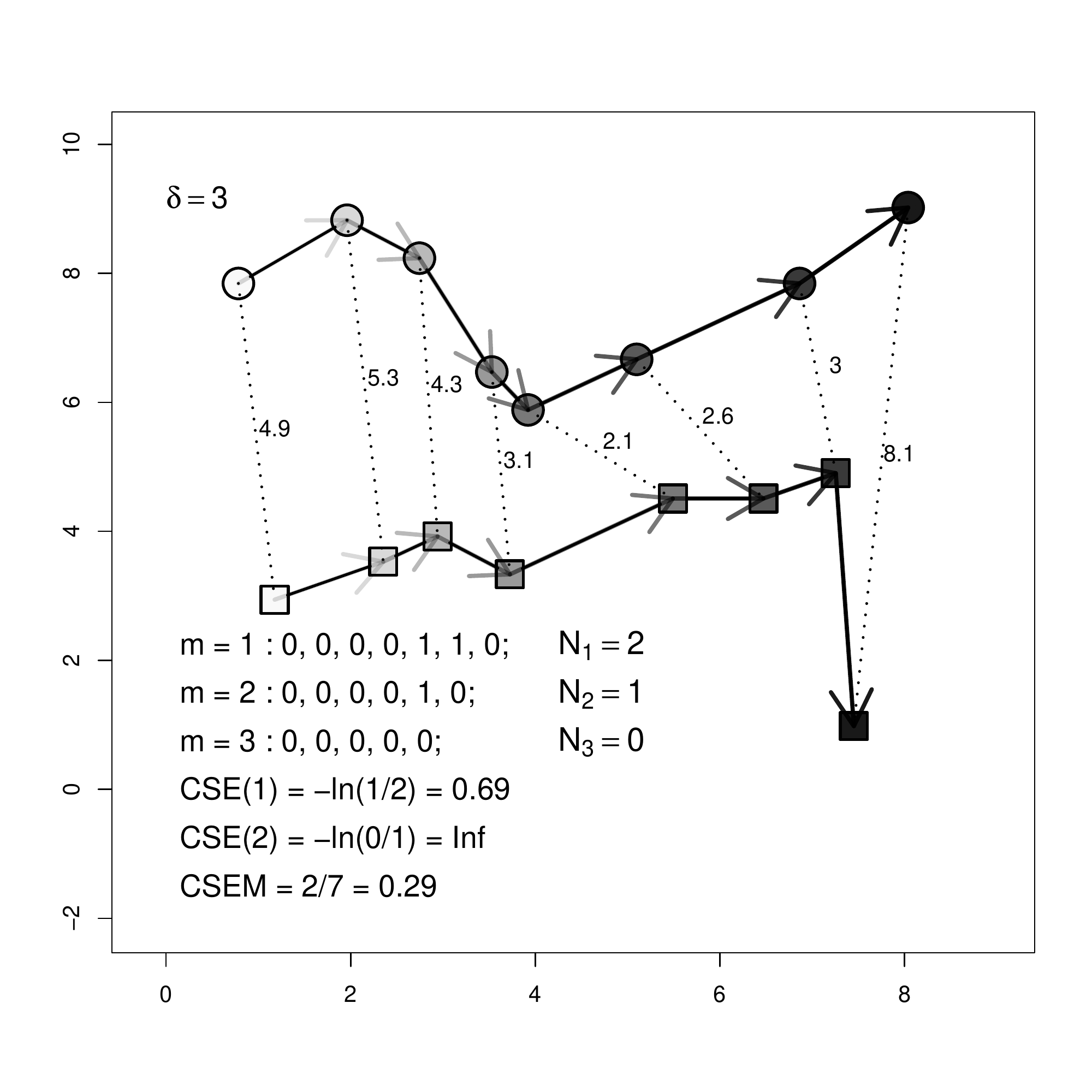}
	\caption{Example of the derivation of CSE and CSEM when the compared features correspond to the positions of the individuals and $\delta = 3$. Circles and squares represent positions of two different individuals (A and B). The grey scales and arrows represent the time sequence of both tracks. Dotted lines represent the distances between simultaneous fixes; their values are shown. Values for all steps for CSEM computation are also shown.}
	\label{CSEEx}
\end{figure}

CSE relies on an ad hoc choice of both $m$ and $\delta$. 
In practice, it is expected that the movement series of A and B will not be constantly synchronous and that, for a large value of $m$, $N_m$ could be equal to $0$, in which case CSE would tend to $\infty$. Therefore, the largest value of $m$ such that $N_m>0$, i.e. the length of the longest similar segment, could be an alternative indicator of similarity between the series (do not confuse with the longest common subsequence LCSS; see \cite{Vlachos2002b}). We propose to use this measure (standardized by $T-1$  to get a value between 0 and 1) as an alternative index of joint movement (formula \ref{CSEMEq}), which we denote by CSEM.  
An example of a dyad and the computation of its CSEs and CSEM is shown in Fig. \ref{CSEEx}.

\begin{equation}
CSEM = \frac{\max \left\{m; N_m  >0\right\} }{T-1}, 
\label{CSEMEq}
\end{equation}
with the convention that $\max{\left\{ \emptyset\right\}}=0$.

\subsection{Correlations ($r_V$)} Pearson and Spearman correlations between variables such as longitude, latitude, distance, velocity, acceleration and turning angles from pairs of tracks, have been used as measures of synchrony in several studies \cite[e.g.][]{Dodge2009}. Correlations are easy to interpret. Pearson correlation coefficients (Table \ref{IndReview}) assess linear correlations, while Spearman correlation coefficients based on ranks statistics capture any functional correlation. The correlation in a given V variable between dyads is denoted by $r_V$.

\subsection{Dynamic Interaction (DI, $\mathbf{DI_d}$ and $\mathbf{DI_\theta}$)} \cite{Long2013} argued that it is necessary to separate movement patterns into direction and displacement (i.e. distance between consecutive fixes or step length), instead of computing a correlation of locations \cite{Shirabe2006} which may carry a mixed effect of both components. To measure interaction in displacement, at each time step, the displacements of simultaneous fixes are compared (formula \ref{DIdEq}).

\begin{equation}
g_t^{\beta} = 1- \left(\frac{\mid d_{t,t+1}^{A} - d_{t,t+1}^{B} \mid}{d_{t,t+1}^{A} + d_{t,t+1}^{B}}\right)^\beta
\label{DIdEq}
\end{equation}	
where $\beta$ is a scaling parameter meant to give more or less weight to similarity in displacement when accounting for dynamic interaction. As $\beta$ increases, $g_t^{\beta}$ is less sensitive to larger differences in displacement. Its default value is $1$. When $d_{t,t+1}^{A} = d_{t,t+1}^{B}$, $g_t^{\beta}=1$; and when the difference in displacement between $A$ and $B$ at time $t$ is large, $g_t^{\beta}$ approaches zero. For $g_t^{\beta}$ to be $0$, one (and only one) of the individuals in the dyad should not move; for a sum of $g_t^{\beta}$ to be equal to zero, at every time $t$ one of the two individuals should not move.

Interaction in direction is measured by

\begin{equation}
f_t = \cos(\theta_{t,t+1}^{A} - \theta_{t,t+1}^{B})
\label{DIthetaEq}
\end{equation}		
where $\theta_{t,t+1}$ is the direction of an individual between time $t$ and $t+1$. $f_t$ is equal to $1$ when movement segments have the same orientation, $0$ when they are perpendicular and $-1$ when they go in opposite directions. \\

\begin{figure}[ht!]
	\centering%
	\includegraphics[scale=0.4]{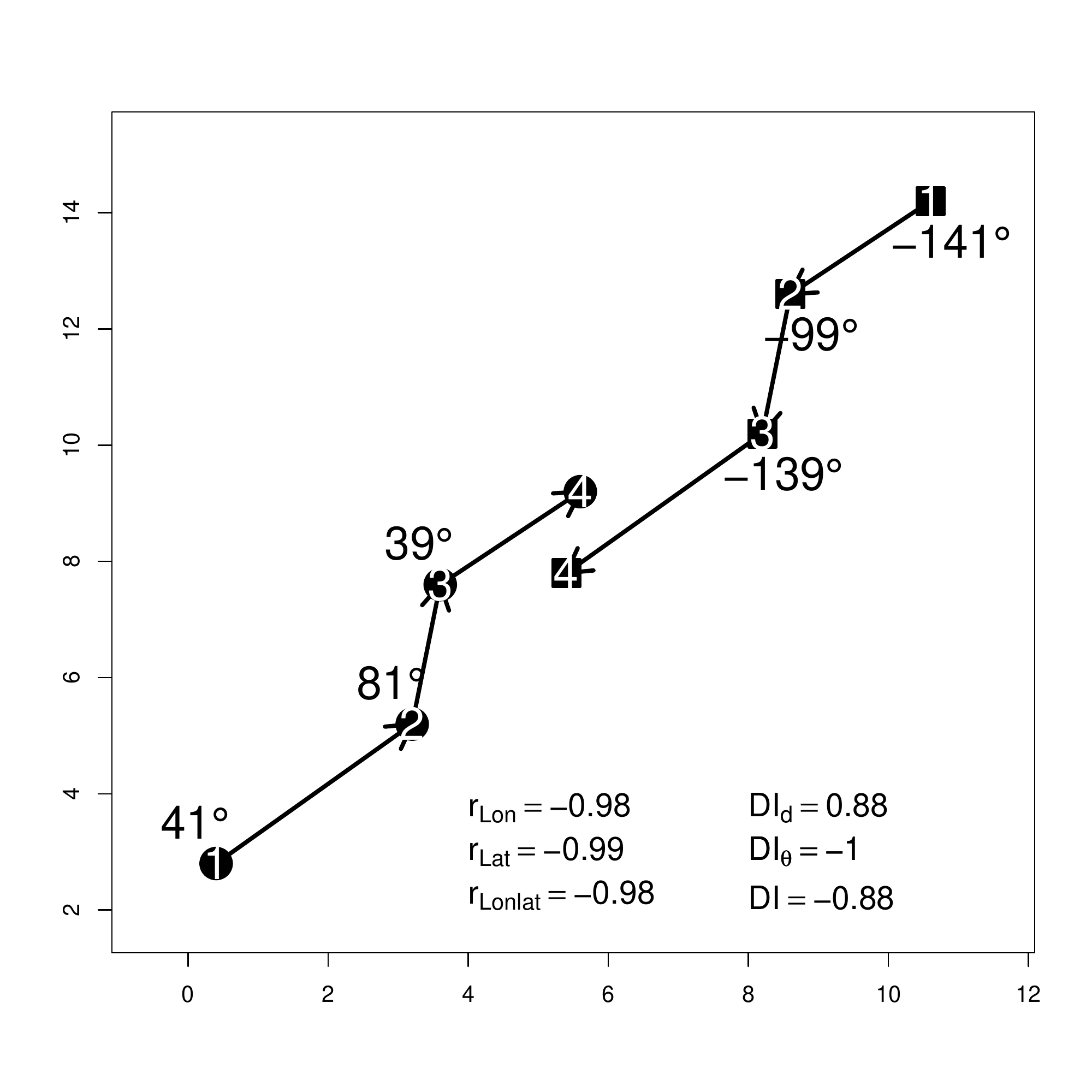}
	\caption{Example of a dyad for which correlations in longitude, latitude and an average of both ($r_{Lon}$, $r_{Lat}$ and $r_{Lonlat}$, respectively), $DI_d$, $DI_\theta$ and $DI$ are derived. Circles and squares represent locations of two different individuals (A and B); the numbers inside represent the time sequence. Displacement lengths and absolute angle values are also shown.}
	\label{DIrEx}
\end{figure}

\cite{Long2013} proposed $3$ indices of dynamic interaction: 1) $DI_d$, dynamic interaction in displacement (average of all $g_t^{\beta}$); 2) $DI_\theta$, dynamic interaction in direction (average of all $f_t$); and 3) DI, overall dynamic interaction, defined as the average of $g_t^{\beta} \times f_t$ (Table \ref{IndReview}). $DI_d$ ranges from 0 to 1, $DI_\theta$ from -1 to 1, and DI from -1 (opposing movement) to 1 (cohesive movement).
Fig. \ref{DIrEx} shows an example of the three indices.

\subsection{Conclusions on the theoretical properties of the metrics}

\paragraph*{Practical use (C1):}
While each metric concerns a concrete aspect of joint-movement behaviour, some of them, such as Cs and DI, are harder to interpret. DI mixes up the coordination in displacement and direction. When DI is close to $1$, it is certainly explained by high values in both components. When it is close to $-1$, it is an indication of overall high displacement coordination but in opposite directions. With values around zero, however, it is impossible to know if it is because of displacement or direction or both. For Cs, because obtaining values close to $-1$ is extremely rare, values around zero and, more particularly, slightly negative values are difficult to interpret. In addition, the maximum attainable value depends on the length of the series, which is likely to vary from dyad to dyad (appendix \ref{appendix:Cs1Negative}). 

\paragraph*{Dependence on parameters (C2):}
Almost every metric depends on the ad hoc definition of a parameter or component, as summarized in Table \ref{IndReview}. This is consistent with the fact that, since there is no consensus on the definition of behaviour \cite{Levitis2009}, and much less on that of collective behaviour, its study depends heavily on the definition that the researcher gives to it. It should be noted that behind each choice of a parameter value, there is also an underlying assumption (e.g. that a distance below a $\delta$ value means proximity); the difference is that parameters can be tuned, and a variety of values can be easily tested. HAI and $L_{ixn}T$ make a critical assumption of a static reference area, and its definition, which may be tricky for highly mobile individuals, is a key issue for the computation of both metrics. On the other hand, $r_V$ and $DI_\theta$ are the only metrics that do not depend on parameter tuning or assumptions for its derivation; except for the assumptions of correlations being linear, or of linear movement between two successive positions when deriving directions, respectively. 

\section{Exploration of metrics through case scenarios}
\label{Simul}

In this section we used schematic, simple and contrasting case scenarios to evaluate the ability of the metrics to assess joint movement, in terms of proximity and coordination.

To build the case scenarios, we considered three levels of dyad proximity (high, medium and low); coordination was decomposed into two aspects, direction (same, independent and opposite) and speed (same or different). Eighteen case scenarios were thus built, with one example of dyad per scenario (Fig. \ref{SceFig}; metrics in appendix \ref{appendix:IndicesTable}). The dyads for each case scenario were deliberately composed of a small number of fixes \cite[$\sim$ 10 simultaneous fixes, as in][]{Long2013} to facilitate interpretation of the metric values and the graphical representation of the arbitrarily constructed tracks (online access to tracks in appendix \ref{appendix:Codes}). To assess the sensitivity of the metrics to changes in patterns of proximity and coordination, the case scenarios were grouped according to the categories in Table \ref{CaseSceTable}. 

\begin{figure}[ht!]
	\centering
	\includegraphics[scale=0.45]{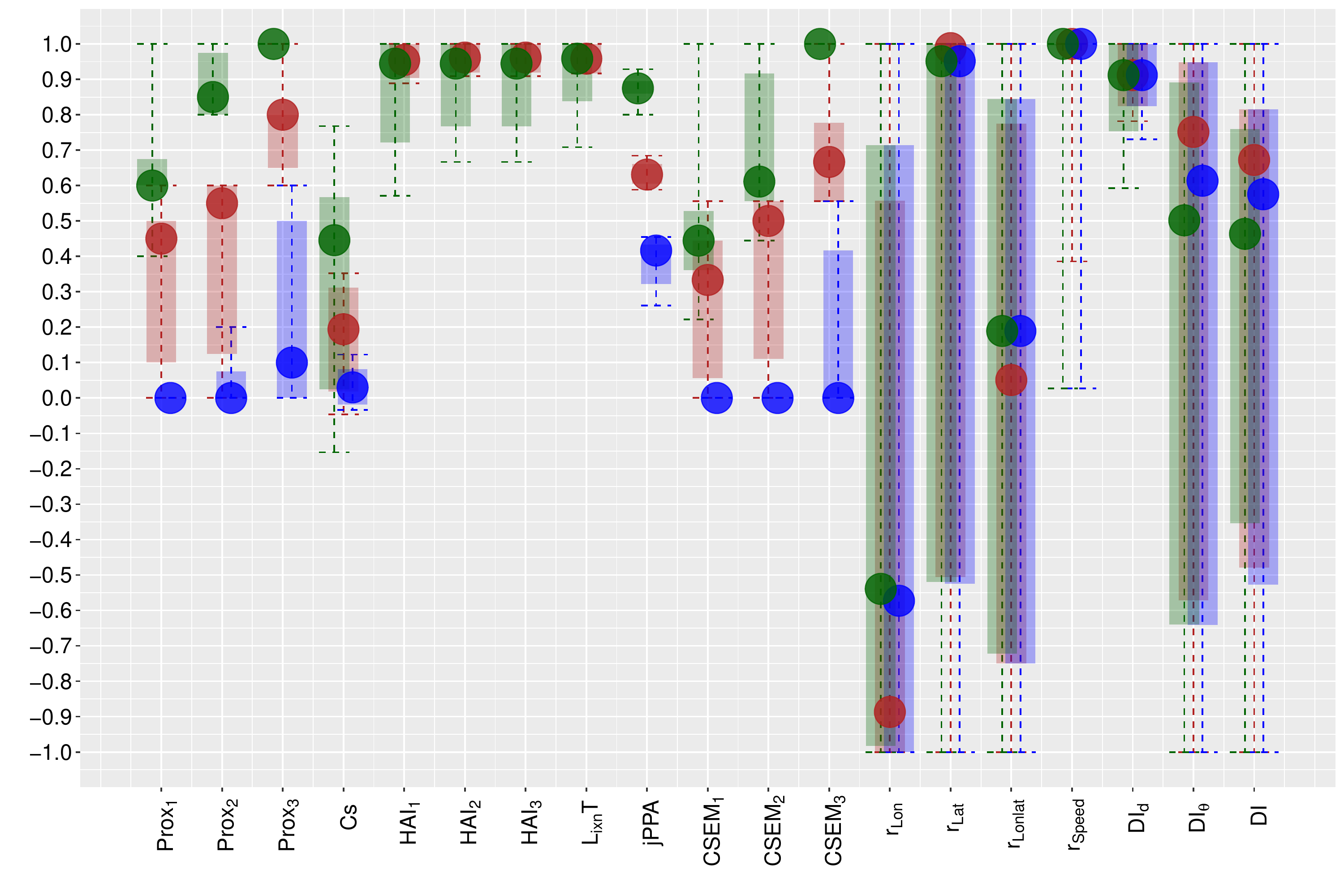}
	\caption{Boxplots of each metric by category of proximity. Green, red and blue correspond to case scenarios of high, medium and low proximity. For each category, the solid circle corresponds to the median, the solid wide bar contains values from the first to the third quartiles, while the dashed line joins the minimum to the maximum values. The green and blue boxplots are shifted to the left and right, respectively, to distinguish them better in case of overlap.}
	\label{ProxSensit}
\end{figure}

\begin{figure}[ht!]
	\centering
	\includegraphics[scale=0.45]{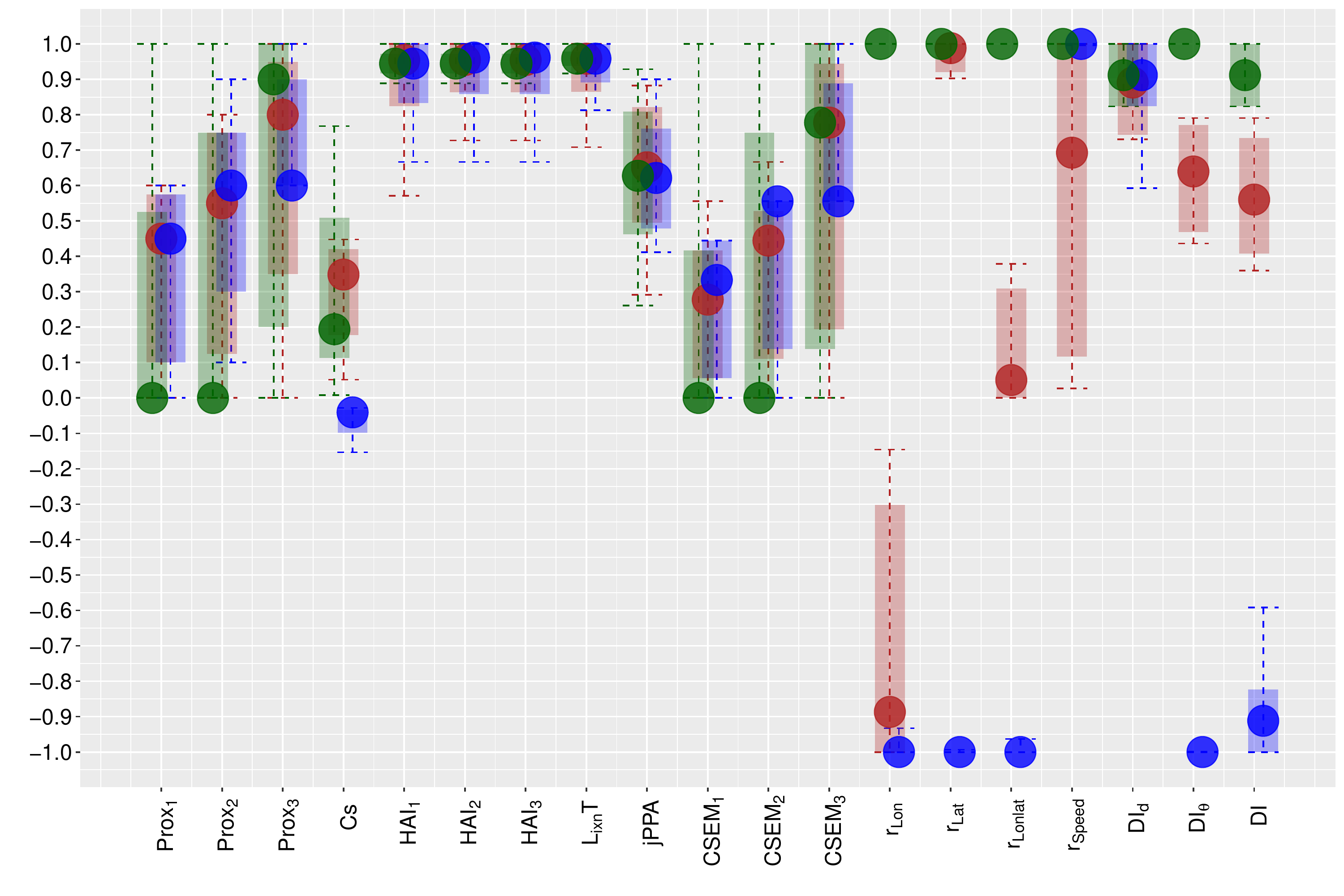}
	\caption{Boxplot of each metric by category of direction coordination. Green, red and blue correspond to case scenarios of same, independent and opposite direction. For each category, the solid circle corresponds to the median, the solid wide bar contains values from the first to the third quartiles, while the dashed line joins the minimum to the maximum values. The green and blue boxplots are shifted to the left and right, respectively, to distinguish them better in case of overlap.}
	\label{DirectionSensit}
\end{figure}

\begin{figure}[ht!]
	\centering
	\includegraphics[scale=0.45]{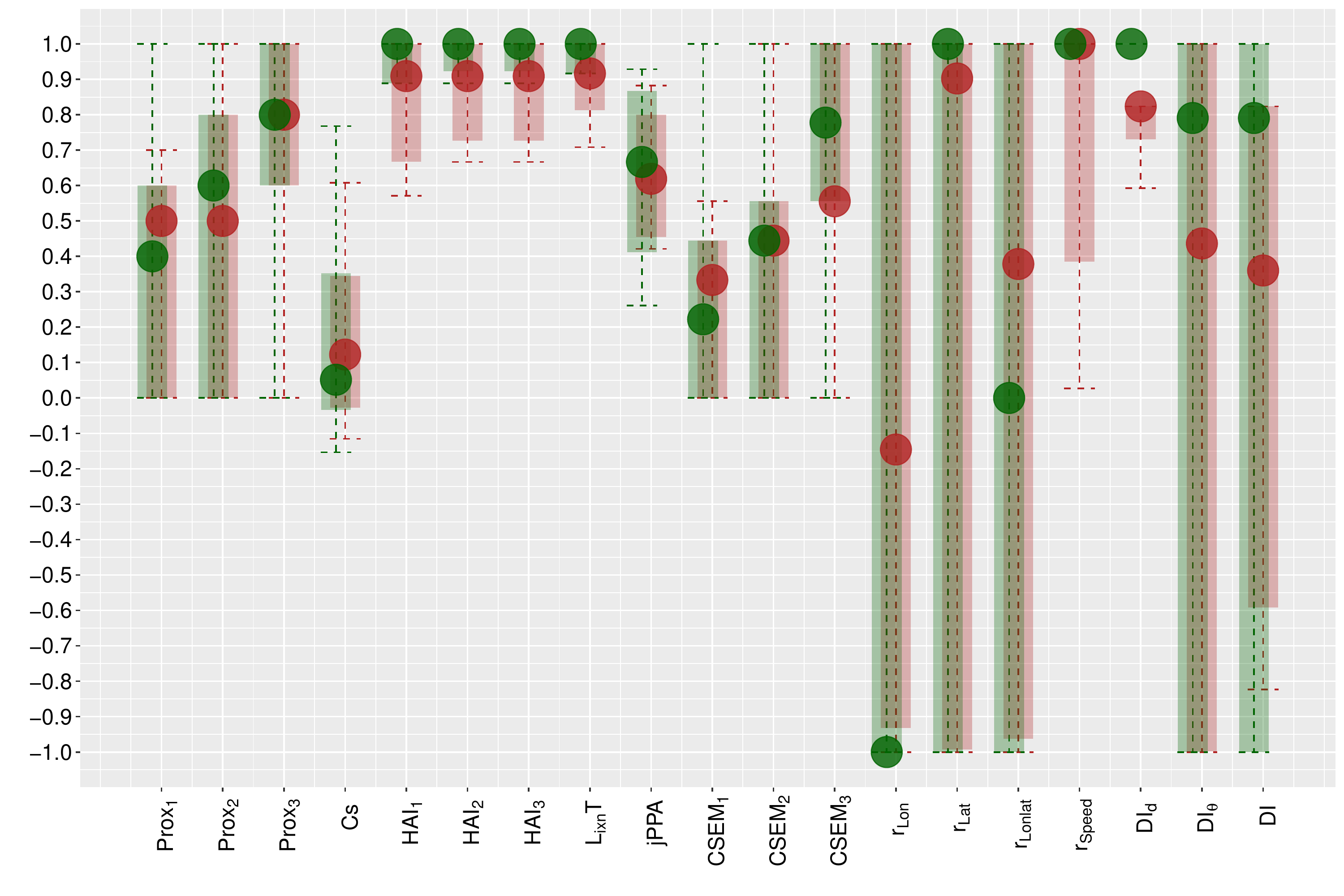}
	\caption{Boxplot of each metric by category of speed coordination. Green and red correspond to case scenarios of same and different speed. For each category, the solid circle corresponds to the median, the solid wide bar contains values from the first to the third quartiles, while the dashed line joins the minimum to the maximum values. The green boxplots are shifted to the left to distinguish them better in case of overlap}
	\label{SpeedSensit}
\end{figure}

Due to the simplicity for its interpretation, Prox was defined as $\Proxdelta$. Three distance thresholds $\Proxdelta$ of 1, 2 and 3 distance units were used for Prox, HAI and CSEM, thus denoted for instance Prox$_1$, Prox$_2$ and Prox$_3$. For Cs, the original definition (equation \ref{CsEq}) was used. jPPA, $\phi$ was arbitrarily fixed to 10. Regarding dynamic interaction, $\beta$ was fixed to 1. 
The $v$ variables for Pearson correlations (Table \ref{IndReview}) were longitude ($r_{Lon}$), latitude ($r_{Lat}$) and speed ($r_{Speed}$). An average of correlations in longitude and latitude, denoted by $r_{Lonlat}$, was also computed. Boxplots of each metric were derived for each proximity and coordination category (Fig. \ref{ProxSensit}, \ref{DirectionSensit} and \ref{SpeedSensit}). \\

\begin{table}[ht!]
	\caption{Case scenarios.}
	\begin{center}
		\begin{tabular}{c|lll}
			\hline
			\hline
			\rule{0pt}{11pt}
			\multirow{2}{*}{ID} & \multirow{2}{*}{Proximity} & \multicolumn{2}{c}{Coordination} \\
            & & Direction & Speed \\
			\hline 
			\rule{0pt}{11pt}
			$1$ & High & Same & Same \\
			$2$ & High & Same & Different \\			
			$3$ & High & Independent & Same \\
			$4$ & High & Independent & Different \\			
            $5$ & High & Opposite & Same \\			
            $6$ & High & Opposite & Different \\			
			$7$ & Medium & Same & Same \\
			$8$ & Medium & Same & Different \\			
			$9$ & Medium & Independent & Same \\
			$10$ & Medium & Independent & Different \\			
            $11$ & Medium & Opposite & Same \\			
            $12$ & Medium & Opposite & Different \\			
			$13$ & Low & Same & Same \\
			$14$ & Low & Same & Different \\			
			$15$ & Low & Independent & Same \\
			$16$ & Low & Independent & Different \\			
            $17$ & Low & Opposite & Same \\			
            $18$ & Low & Opposite & Different \\			
			\hline
			\hline
		\end{tabular}
	\end{center}
	\label{CaseSceTable}
\end{table} 

\begin{figure}[ht!]
	\centering
	\includegraphics[scale=0.5]{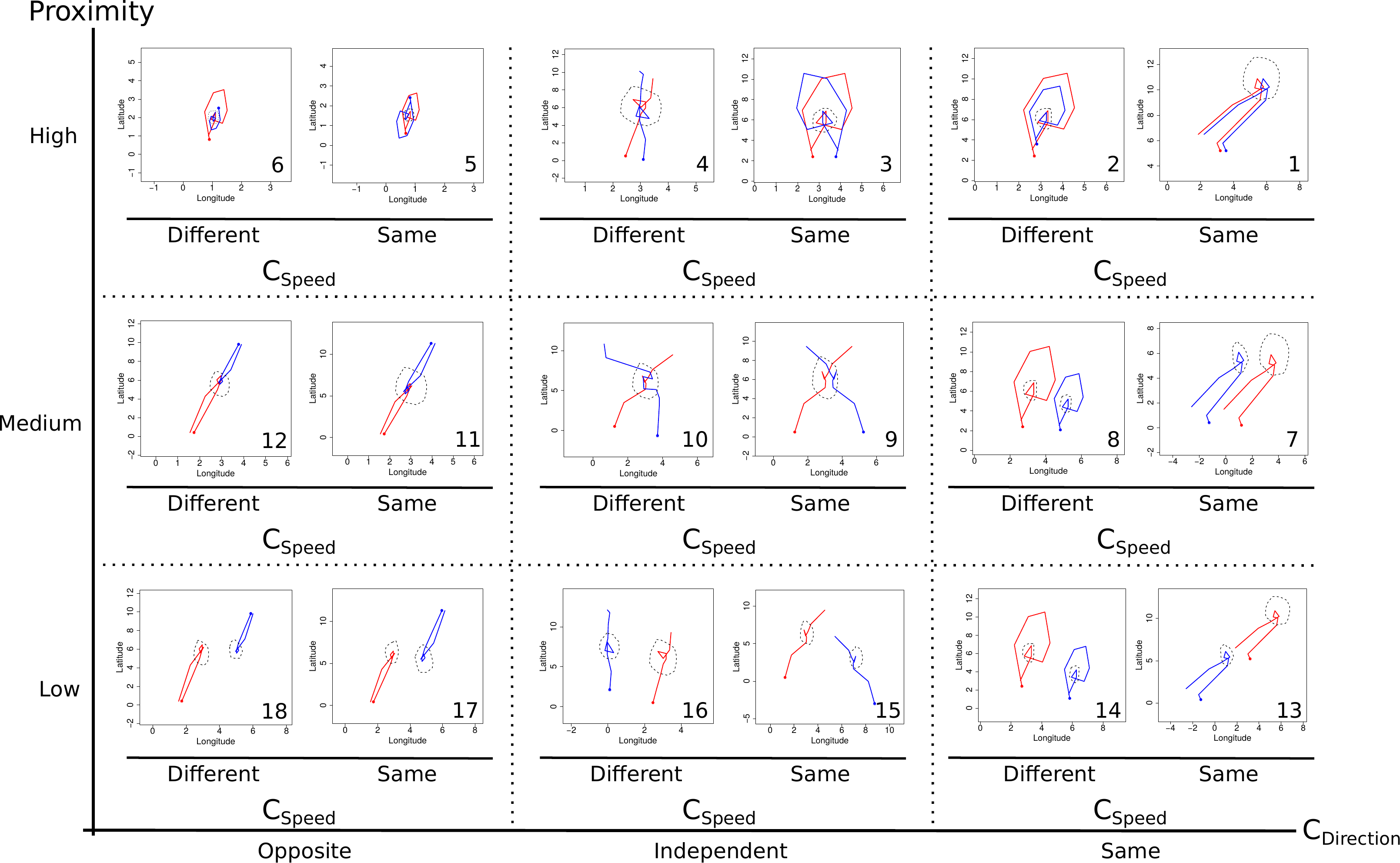}
	\caption{One example of dyad for each case scenario representing contrasting patterns of proximity and coordination (in direction and speed, $C_{Direction}$ and $C_{Speed}$, respectively); numbers correspond to scenario ID in Table \ref{CaseSceTable}. Solid lines represent the two trajectories, the solid points correspond to the start of the trajectories. The black dashed circumferences represent arbitrary reference areas; two circumferences correspond to an absence of a common reference area.}
	\label{SceFig}
\end{figure}

The values taken by Prox, jPPA, CSEM and, to a lesser degree, Cs, showed sensitivity to the level of proximity (Fig. \ref{ProxSensit}). Conversely, no association was revealed between the proximity scenarios and the metrics based on correlation, dynamic interaction and reference area occupation. 

Changes in direction were reflected in values taken by correlation metrics on location ($r_{Lonlat}$, $r_{Lon}$ and $r_{Lat}$) and two dynamic interaction metrics, DI and $DI_\theta$ (Fig. \ref{DirectionSensit}). Cs took lower values in scenarios of opposite direction, but independent and same direction scenarios reflected no distinction for this metric. High correlation in speed was found for scenarios of opposite and same direction, while a large variability was found when direction was independent. $r_{speed}$ showed differences when direction was independent between dyads, but no distinction was caught by the metric between same and opposite direction scenarios. The other metrics did not show distinguishable patterns related to changes in direction coordination. 

Concerning coordination in speed, the most sensitive metric was $DI_d$, which measures similarity in the distances covered by individuals at simultaneous fixes (Fig. \ref{SpeedSensit}). $r_{Speed}$ took a wide range of values when speed was not coordinated, while it was equal to 1 when perfectly coordinated. $DI_d$ is more sensitive to changes in the values of speed (similar to step length because of the regular step units) than $r_{speed}$ which characterizes variations in the same sense (correlation), rather than correspondence in values. HAI and $L_{ixn}T$ showed slight differences in their ranges of values with changes in speed-coordination scenarios. When analysing combined categories of proximity and speed-coordination, and proximity and direction-coordination, less distinctive patterns were found, probably due to the higher number of categories, each containing fewer observations (Figs. in Appendix \ref{appendix:IndicesFigs}).

Overall, Prox, jPPA, CSEM, $r_{Lonlat}$, $r_{Speed}$, $DI_d$, $DI_\theta$ and DI were highly sensitive to changes in patterns of either proximity or coordination. For proximity scenarios, the variance of some metrics for each category was also sensitive to the $\delta$ chosen; i.e. for larger $\delta$, the variance of Prox and CSEM decreased in high proximity, while it increased for low proximity cases. This pattern does not hold for HAI, probably due to the strong dependence of this metric on the arbitrary choice of the reference area. Cs showed a slight sensitivity to changes in direction and proximity scenarios, although the values taken for each type of case scenario did not show a clear separation. 

\section{Assessment of computational cost}
\label{CPUtime}

For the estimation of the computational cost, we needed larger series and numerous dyads. Therefore, we simulated $1000$ dyads with trajectories following a Brownian motion, each one composed of $100$ fixes. Table \ref{CompTable} shows the summary statistics for the CPU time for computing each metric for one dyad. In all cases, it took less than a second per dyad; but the small differences may become relevant for large datasets. jPPA took the highest CPU time; more than 50 times higher than the second highest one, i.e. CSEM. It should be noted that for jPPA, the areas of intersection and union of the ellipses were approximated by grid cells, so for smaller cell sizes (i.e. more accurate jPPA estimation), the computational cost would increase. Because individuals were given freedom to go in any direction, no assumption on a reference area was made, thus $L_{ixn}T$ and HAI were not considered in this part of the analysis. It is clear from their mathematical definition in Table \ref{IndReview} that their CPU time would be higher than the one of Prox but lower than the one of CSEM.

\begin{table}[ht!]
	\caption{Median CPU time for computing each metric for a given dyad (in $10^{-4}$ seconds); the 2.5 and 97.5 percentiles are in brackets.}
	\begin{center}
		\begin{tabular}{ll}
			\hline
			\hline
			\rule{0pt}{11pt}
			Metric & Time \\
			\hline 
			\rule{0pt}{11pt}
			$Prox$ & $7.32$ $[6.85,12.21]$ \\
			$Cs$ & $13.52$ $[12.36,60.99]$ \\			
			$jPPA$ & $5171.36$ $[4917.48,6990.36]$ \\			
			$r_{Lonlat}$ & $1.34$ $[1.12,1.96]$ \\
			$r_{Speed}$ & $1.29$ $[1.12,1.88]$ \\
			$CSEM$ & $87.85$ $[29.89,196.42]$ \\
			$DI$ & $2.32$ $[2.10,3.09]$ \\
			$DI_d$ & $1.52$ $[1.35,2.11]$ \\
			$DI_{\theta}$ & $2.03$ $[1.82,2.75]$ \\
			\hline
			\hline
		\end{tabular}
	\end{center}
	\label{CompTable}
\end{table} 


\section{Synthesis of metric analysis}

Table \ref{IndReviewFinal} summarizes the theoretical and case-scenario analyses. Most metrics reflected marked properties of dyadic joint movement, evidenced both theoretically and through the case scenario assessment. Exceptions were Cs, HAI and $L_{ixn}T$. Cs was sensitive to the null model for the distance expected by chance ($D_{chance}$; formula \ref{CsEqGen}), it did not attain its whole range of definition, turned out to be asymmetric and dependent on the length of the series (App. \ref{appendix:Cs1Negative}), and was less sensitive than the other metrics to changes in patterns of joint movement. Perhaps a change in the null model for $D_{chance}$ could improve Cs's power to assess joint movement, though the new null model should be justified. HAI and $L_{ixn}T$, dependent on the reference area definition, were even less sensitive to changes in joint movement patterns. This supports our earlier statement that $L_{ixn}T$ and HAI should only be used when a reference area exists and is known. Alternatively, Prox works as a simpler metric and is highly sensitive to changes in proximity. The only drawback of Prox is the need to choose a distance threshold parameter, eventually based on prior knowledge of the spatial dynamics of the population. Otherwise, a set of values can be tested, as shown here. jPPA presents the advantage of not requiring the knowledge of a reference area, but still relies on assumptions related to equal probability of presence in an ellipse, which strongly depends on a $\phi$ parameter whose tuning is not obvious. The use of dynamically changing $\phi$ parameters for jPPA \cite{Long2015d} should be further investigated, but that would likely increase the computational cost of the metric, which already takes more than 50 times the CPU time of the second most expensive metric, CSEM. 

CSEM evaluates the similarity between the dynamical changes in movement patterns within a $\delta$ bandwidth, and, because of that, was expected to be more sensitive to changes in proximity than in coordination. It should be further assessed if using other variables for deriving CSEM (i.e. using \cite{Richman2000} generic definition) could make it more sensitive to coordination than proximity. As with Prox, it is in the hands of the user to tune the threshold parameter. Because we were using locations as the analysed series (so the dynamical changes assessed were in fact changes in distance), we used exactly the same threshold values as for Prox. By contrast, correlations in location ($r_{Lon}$, $r_{Lat}$, $r_{Lonlat}$) did show sensitivity to changes in coordination, as expected. The same occurred with $DI_\theta$ and DI. Correlation in speed was sensitive to changes in both coordination components, showing high variance when there was no coordination (independent direction or speed). $DI_d$, on the other hand, was only sensitive to changes in speed. Because the time-step was regular, identical speed was equivalent to identical covered distance (at simultaneous fixes), which explained how in those scenarios $DI_d$ was equal to $1$. While DI behaved more similarly to $DI_\theta$, its definition makes it impossible to separate the effects of coordination in displacement and in azimuth, which makes the interpretation of the metric more difficult than interpreting $DI_d$ and $DI_\theta$ independently.

Several works discuss the importance of scale and granularity in the analysis of movement patterns \cite{Laube2007,Laube2011,DeSolla1999}. For this study we made the implicit assumption that granularity was right for the case scenarios, but this would be an issue to take into account with real data and is case-specific. Here, the only metric taking into account several scales for analysing joint movement (in terms of similarity and closeness) was CSEM. 

\begin{table}[H]
		\caption{Evaluation of the three criteria for each metric}
	{
    \renewcommand{\baselinestretch}{1.1} 
	\small
		\begin{center}
			\begin{tabular}{>{\centering\arraybackslash}p{1.2cm}p{1.3cm}p{3cm}p{1.4cm}p{1.4cm}p{1.4cm}p{2.8cm}p{1.4cm}}
				\hline
				\hline
				\rule{0pt}{11pt}
                \multirow{4}*{Metric} &  \multicolumn{7}{c}{Criterion} \\
\cmidrule{2-8}
				&  \multicolumn{5}{c}{C1: Practical use} & {C2: Dependence} & {C3: } \\
                \cmidrule{2-6}
                         & Attainable range & Interpretation for joint movement & \multicolumn{3}{c}{Sensitivity to} & on parameters / assumptions & Comp. cost \\
                         \cmidrule{4-6}
                         & & & P & $C_{Direction}$ & $C_{Speed}$ & & \\
				\hline 
\rowcolor{gray!50}				$Prox$ &   \textbf{Yes} & From always distant (0) to always close (1) & \textbf{High} & Low & Low & \textbf{User tractable} (ad hoc definition of distance threshold) & Low \\
				
				$Cs$ & No & Difficult: i) negative value close to 0 difficult to interpret; ii) series-length dependent & Medium & Medium & Low & Not user tractable (null hypothesis of independent movement) & \textbf{Low} \\
			$HAI$  & \textbf{Yes} & From always distant and out of $S_{AB}$ at least for one individual (0) to always close and in $S_{AB}$ (1) & Low & Low & Medium & Not user tractable (reference area and distance threshold) & -- \\
		   $L_{ixn}T$ & \textbf{Yes} & Same as $HAI$ & Low & Low & Medium & Not user tractable (reference area) & -- \\
\rowcolor{gray!25}        $jPPA $ & \textbf{Yes} & From no (0) to permanent (1) potential overlap &  \textbf{High} & Low & Low & \textbf{User tractable} (maximum velocity) & High \\
\rowcolor{gray!50}	    $CSEM $ & \textbf{Yes} & From highly synchronous (0) to asynchronous (1) & \textbf{High} & Low & Low & \textbf{User tractable} (distance threshold) & Medium \\
 \rowcolor{gray!50}       $r_V$ & \textbf{Yes} & From anticorrelated (-1) to correlated (1) & Low & \textbf{High*} & \textbf{High*} & No dependence & \textbf{Low} \\	
\rowcolor{gray!50} 		$DI_d$ & \textbf{Yes} & From opposite (-1) to cohesive (1) movement in displacement & Low & Low & \textbf{High} & User tractable (weighting coefficient for similarity in displacement) & \textbf{Low} \\
\rowcolor{gray!10}          $DI_\theta$ & \textbf{Yes} & From opposite (-1) to cohesive (1) movement in azimuth & Low & \textbf{High} & Low & \textbf{No dependence} & \textbf{Low} \\
\rowcolor{gray!10}        $DI$ & \textbf{Yes} & From opposite (-1) to cohesive (1) movement in both mixed displacement and azimuth effects & Low & \textbf{High} & Low & \textbf{User tractable} (weighting coefficient for similarity in displacement) & \textbf{Low} \\
				\hline
				\hline
			\end{tabular}
		\end{center}
        }
		\begin{tablenotes}
			\small
			\item \textit{Note:} P $=$Proximity, $C_{speed} =$ coordination in speed, $C_{direction} =$ coordination in direction, S $=$ reference area. *Depending on $v$ (see section \ref{Simul}). Text in bold correspond to positive attributes.
		\end{tablenotes}
		\label{IndReviewFinal}
	\end{table}

\begin{figure}[ht!]
	\centering
	\includegraphics[scale=0.75]{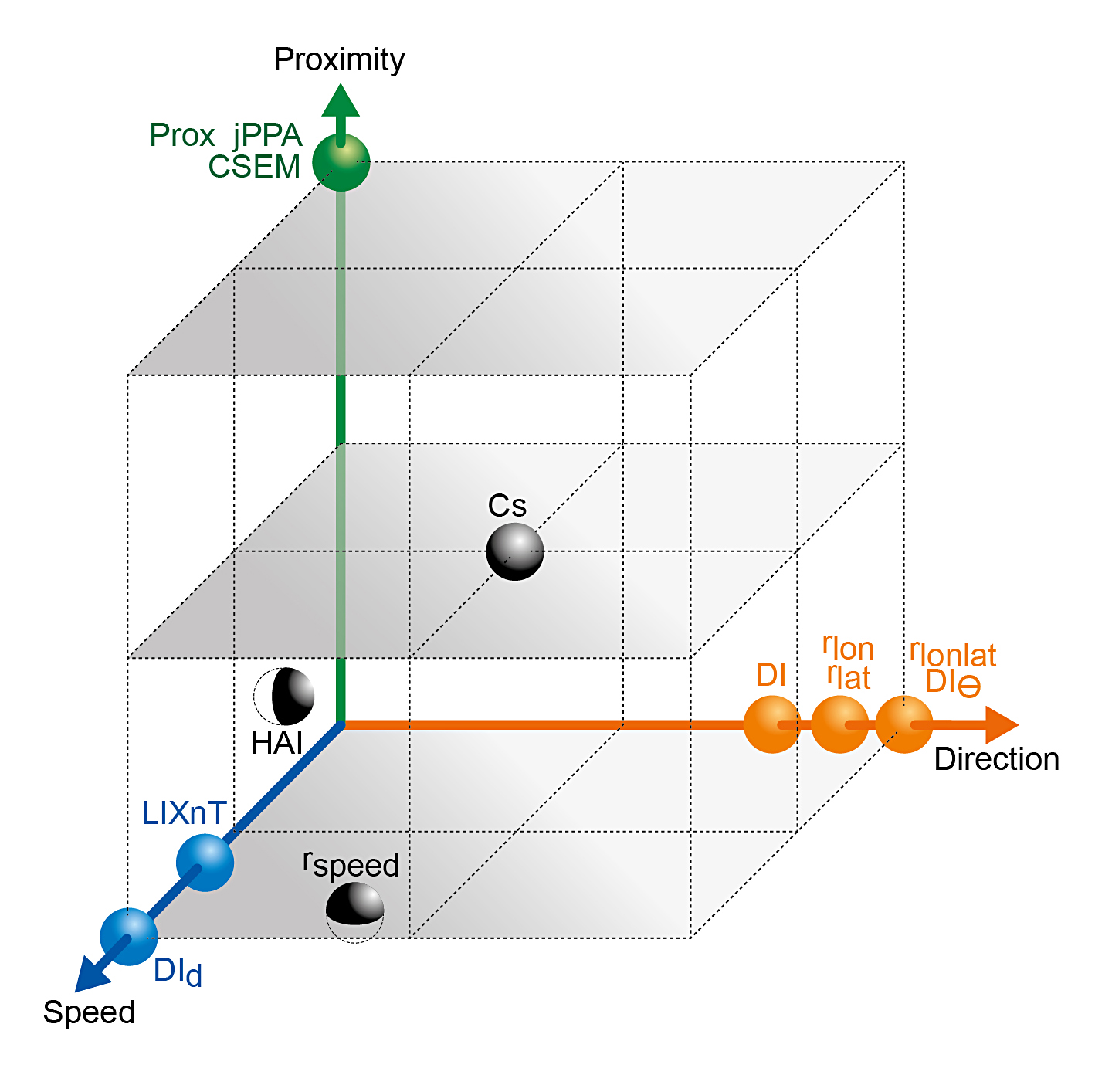}
	\caption{Representation of metrics in terms of their distance relative to proximity and coordination.}
	\label{IndRule}
\end{figure}

We expected to obtain a binary classification of the metrics into proximity and coordination, based on the theoretical and case scenario evaluations. This was not so straightforward and we ended up instead with a 3-dimensional space representation (Fig. \ref{IndRule}). Prox and CSEM are the most proximity-like indices. jPPA would be the third one due to its sensitivity to changes in proximity in the case scenario evaluation. Cs would be somewhere between Prox and direction coordination because it showed certain sensitivity to both HAI and $L_{ixn}T$ are almost at the origin but slightly related to speed coordination. Theoretically, both metrics should account for proximity, since when two individuals are together in the same area, they are expected to be at a relative proximity; in practice, this was not reflected in sensitivity to proximity from HAI and $L_{ixn}T$. Still, HAI is represented in the graphic slightly above $L_{ixn}T$ since its formulation specifically accounts for proximity in solitary use of the reference area. They are both graphically represented in association with the speed coordination axis because of the case scenario results which reflected that being in the same area only simultaneously requires some degree of synchrony. $DI_d$ was the most sensitive metric to speed coordination, followed by $r_{Speed}$. $DI_{\theta}$ and $r_{Lonlat}$ are the most strongly linked to direction coordination, seconded by DI, which is also related to speed coordination. A principal component analysis (PCA) using the values obtained for the case scenarios gave very similar results to those in Fig. \ref{IndRule} (appendix \ref{appendix:PCAresults}), but this schematic representation is more complete because: 1) the theoretical and case-scenario assessment were both taken into account; 2) the PCA was performed without $L_{ixn}T$ and HAI that had missing values for case scenarios with no common reference area (data imputation as in \cite{Josse2009} was not appropriate for this case).

Figure \ref{IndRule} and Table \ref{IndReviewFinal} could be used as guidelines to choose the right metrics depending on the user's case study. For instance, in an African lion joint-movement study \cite{Benhamou2014}, proximity was the focus of the study; in that case, the $I_{AB}$ (Prox) metric was used. For similar studies several proximity-related metrics could be chosen; the choice would depend on the assumptions that the researcher is willing to make. In other cases, researchers may want to assess collective behaviour in tagged animals (e.g. birds or marine mammals) that do not remain proximal during their foraging/migration trips. Then, the collective behaviour component that could be evaluated would be coordination. Whether it is in direction or speed would depend on the researcher's hypotheses. Coordination, or synchrony, has already been observed in some animal species such as northern elephant seals  \cite[e.g.][]{Duarte2018} and bottlenecked sea turtles \cite[e.g.][]{Plot2012}, among others. The use of the metrics presented here would allow a quantification of the pairwise behavioural patterns observed, a first step towards a quantitative analysis of the factors explaining those behaviours (e.g. physiological traits, personality or environmental conditions). The metrics presented here are applicable to any organism with tracking data (not necessarily georeferenced). 

If the aim is to evaluate all three joint-movement dimensions, we advice to consider for each dimension at least one metric that is highly sensitive to it, rather than a metric that is weakly related to two or three. The complementarity of the metrics (i.e. multivariate approach) has not been studied here, and should be the focus of a future study.

\section{Further perspectives on collective behaviour}

The assessment of a `lagged-follower' behaviour, where one individual would follow the other, was out of the scope of this work and should be addressed in the future. The study of this type of interactions is rather challenging, since the lag in the following behaviour is probably not static, and could vary between tracks and also within tracks. Some recent works have focused on this type of interaction regarding it as a delay between trajectories, which transforms the problem into one of similarity between trajectories, where one is delayed from the other \cite{Giuggioli2015,Konzack2017}. Metrics based on the Fr\'echet distance \cite{Frechet1905,Aronov2006} or the Edit distance \cite{Levenshtein1966} are common choices for measuring those similarities in computer science studies. In terms of computational cost, assessing following behaviour should be much more expensive than assessing joint movement.

This study focused on dyadic joint movement. The next step would be to identify metrics to characterize collective behaviour with more than two individuals. A pragmatical approach to investigate this more complex issue could be to identify, within large groups of individuals, the ones that move together for each given segment of trajectories (as dyads, triads or larger groups), and to study those dynamics. A similar procedure could then be used to spot following behaviour and leadership. Movement could be then regarded as spatio-temporal sequences of joint, following, hybrid and independence movement with one or more partners. 

Finally, a robust assessment of the different patterns of collective behaviour (e.g. proximal joint movement, coordination movement, follower movement) at multiple scales would provide realistic inputs for including group dynamic into movement models, which until now have relied on strong assumptions on collective behaviour in the few cases where it was taken into account \cite{Langrock2014,Haydon2008,Potts2014,Niu2016,Russell2016}, mostly due to the lack of understanding of collective motion. 

\section{Acknowledgements} This work has received funding from French region Pays de la Loire and the research project COSELMAR, the French research network PathTIS and the European Union’s Horizon 2020 research and innovation program under grant agreement No 633680, Discardless. The authors would like to warmly thank Mathieu Basille for his constructive comments to the manuscript, Angela Blanchard for proofreading and Criscely Lujan for help with github. 



\section{Supporting information and appendices}

\noindent\ref{appendix:ProxDif}. Graphical examples of two kernel functions for Proximity metrics

\noindent\ref{appendix:Cs1Negative}. Cs1 requirements to take large negative values

\noindent\ref{appendix:pLixn}. Lixn: Table for computing probabilities

\noindent\ref{appendix::Ellipse}. How to define the ellipse of the potential path area

\noindent\ref{appendix:IndicesTable}. Metrics derived for each case scenario

\noindent\ref{appendix:IndicesFigs}. Summary figures for proximity-speed and proximity-coordination scenarios

\noindent\ref{appendix:PCAresults}. Principal component analysis of the metrics for the case scenarios.

\noindent\ref{appendix:Codes}. Codes for computing the metrics.

 \bibliographystyle{hapalike.bst}


\appendix
\renewcommand{\thesection}{S\arabic{section}}   

\section{Graphical examples of two kernel functions for Proximity metrics}
\label{appendix:ProxDif}

\setcounter{figure}{0}
\renewcommand{\thefigure}{\ref{appendix:ProxDif}.\arabic{figure}}

\begin{figure}[ht!]
		\centering%
		\includegraphics[scale=0.65]{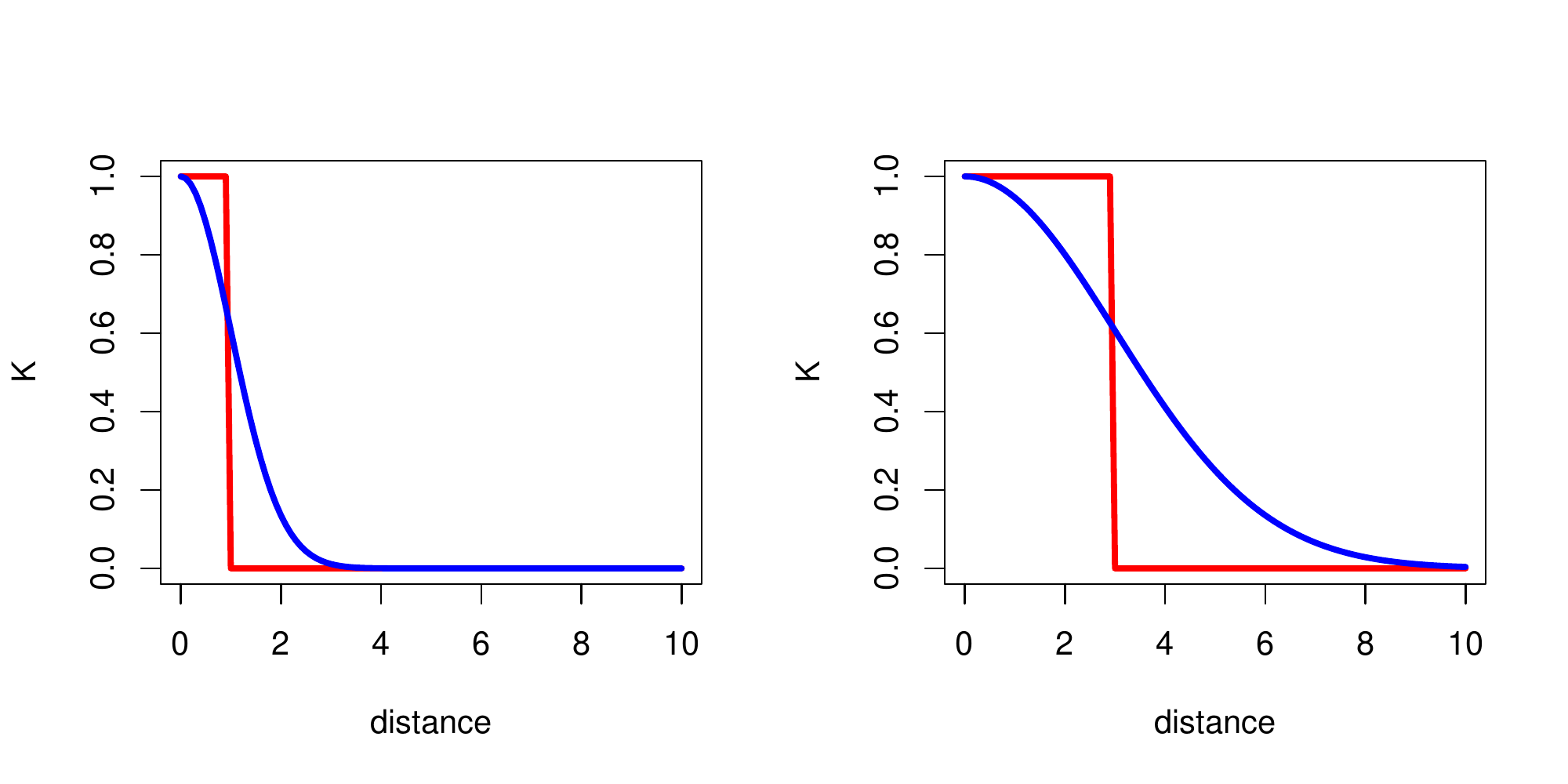}
	\caption{Illustration of differences between to kernel values with the same $\delta$ value. The red and blue solid lines corresponds to $K_{\delta}(x,y) =\mathbbm{1}_{\{ \| x-y \| < \delta \}}$ and $K_{\delta}(x,y)=\exp\big(- \| x-y \|^2  /(2 \delta^2)\big)$, respectively. Left panel: $\delta=1$. Right panel: $\delta=3$.}
	\label{ProxComp}
\end{figure}

Figure \ref{ProxComp} shows smoothed functions when $K_{\delta}(x,y)=\exp\big(- \| x-y \|^2  /(2 \delta^2)\big)$, i.e. $I_{AB}$, but those values are hard to interpret. $K_{\delta}(x,y) =\mathbbm{1}_{\{ \| x-y \| < \delta \}}$ is not continuous, but the interpretation of a change in $\delta$ is straightforward.

\section{Cs1 requirements to take large negative values}
\label{appendix:Cs1Negative}

\setcounter{figure}{0}
\renewcommand{\thefigure}{\ref{appendix:Cs1Negative}.\arabic{figure}}

Let $d_{ij}$ be the distance between the locations of $A$ at time $i$ and $B$ at time $j$. Then, $D_O$ and $D_E$ can be expressed as in equations \ref{Dii} and \ref{De}.

\begin{equation}
D_O = \ds\sum_{i_1=1}^{T}d_{ii}/T
\label{Dii}
\end{equation}

\begin{equation}
D_{\bar{O}} = \ds\sum_{\substack{i,j \in [1,T] \\ i \neq j}}d_{ij}/(T^2-T)
\label{Dij}
\end{equation}

\begin{equation}
D_{E} = \ds\frac{D_O}{T} + \ds\frac{(T-1)}{T}D_{\bar{O}}
\label{De}
\end{equation}

where $D_{\bar{O}}$ is defined in equation \ref{Dij} and corresponds to the average distance between the exclusively permuted points without taking into account the simultaneous fixes. Using those equations, we can replace $D_O$ and $D_E$ in equation \ref{CsEq} when $Cs1=-\alpha$ ($\alpha > 0$) and obtain:

\begin{equation}
\frac{D_{\bar{O}}}{D_O} = \ds\frac{T(1-\alpha)}{(T-1)(1+\alpha)} - \ds\frac{1}{T-1}
\label{dEq}
\end{equation}

\begin{figure}[ht!]
		\centering%
		\includegraphics[scale=0.5]{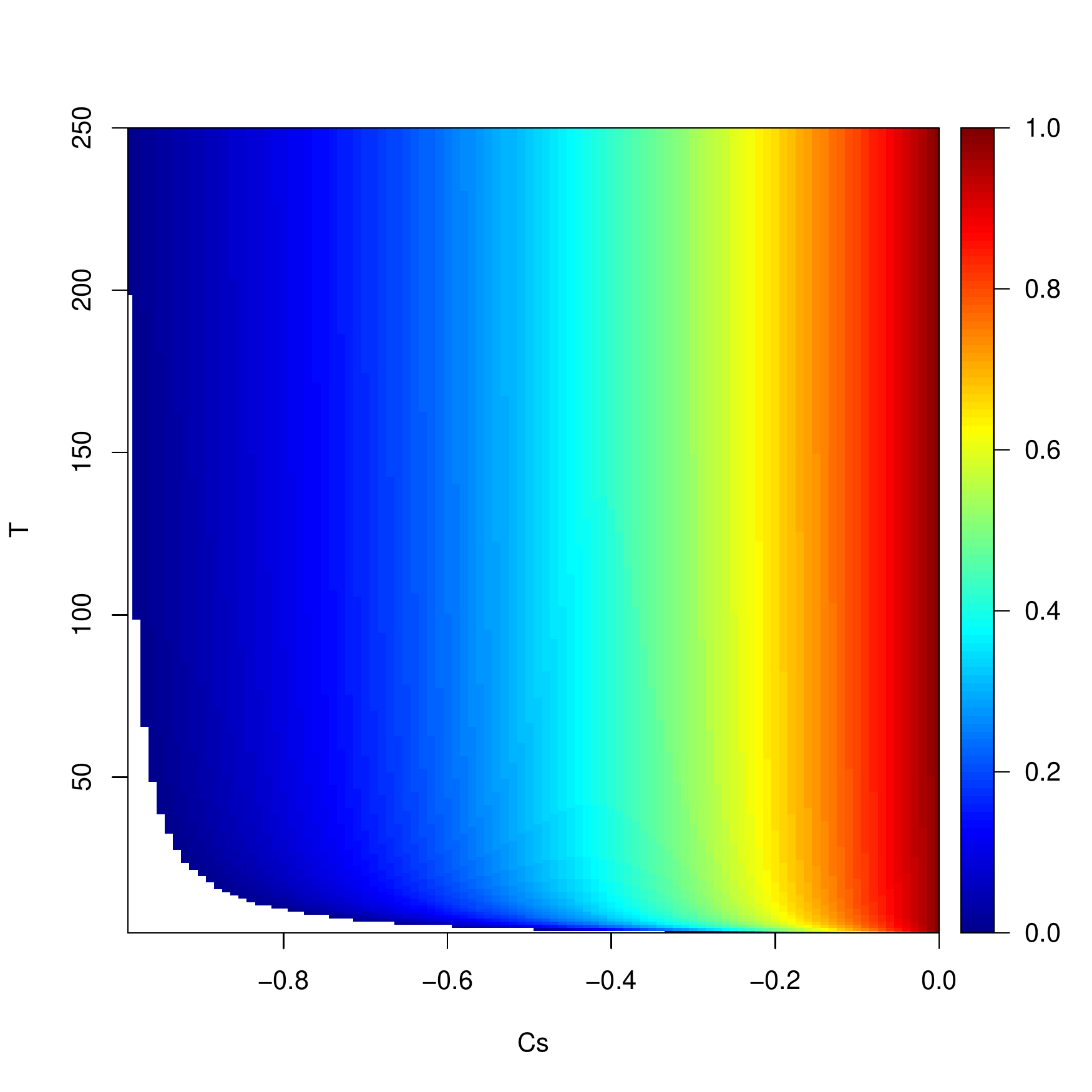}
	\caption{Computed ratios $D_{\bar{O}}/D_{O}$ needed for obtaining the Cs1 negative values (x-axis, from $-0.99$ to $0$) for each series length T (y-axis, from $2$ to $250$). The blank spaces correspond to infeasible situations.}
	\label{CsNeg}
\end{figure}

It means that, for instance, for $Cs1=-0.5$ and when $T$ is large, $D_{\bar{O}}$   would have to be approximately a third of $D_{{O}}$, thus a third of the average distance computed only at simultaneous fixes. Fig. \ref{CsNeg} shows the values of $D_{\bar{O}}/D_{O}$ ratios needed to attain the whole range of Cs negative values. Most of those scenarios are very unlikely. 

\section{Lixn: Table for computing probabilities}
\label{appendix:pLixn}

\setcounter{table}{0}
\renewcommand{\thetable}{\ref{appendix:pLixn}.\arabic{table}}

\begin{table}[H]
	\caption{Probabilities of finding individuals A and B simultaneously or not in the reference areas.}
	\begin{center}
		\begin{tabular}{cll }
			\hline
			\hline
			\rule{0pt}{11pt}
			Probability & With individual areas & With expected frequency \\
			\hline 
			\rule{0pt}{20pt}
			$p_{AB}$ & $\ds\frac{S_{AB}}{S_{A}}\times\ds\frac{S_{AB}}{S_{B}}$ & $\ds\frac{n_A}{T} \times \ds\frac{n_B}{T}$ \\
			\rule{0pt}{20pt}
			$p_{A0}$ & $\ds\frac{S_{AB}}{S_{A}}\times\left(1 - \ds\frac{S_{AB}}{S_{B}}\right)$ & $\ds\frac{n_A}{T} \times \left(1-\ds\frac{n_B}{T}\right)$ \\				
			\rule{0pt}{20pt}
			$p_{0B}$ & $\left(1 - \ds\frac{S_{AB}}{S_{A}}\right)\times\ds\frac{S_{AB}}{S_{B}}$ & $\left(1-\ds\frac{n_A}{T} \right)\times \ds\frac{n_B}{T}$ \\
			\rule{0pt}{20pt}
			$p_{00}$ & $\left(1 - \ds\frac{S_{AB}}{S_{A}}\right)\times\left(1-\ds\frac{S_{AB}}{S_{B}}\right)$ & $\left(1-\ds\frac{n_A}{T} \right)\times \left(1-\ds\frac{n_B}{T}\right)$ \\				
			\hline
			\hline
		\end{tabular}
	\end{center}
	\begin{tablenotes}
		\small
		\item \textit{Note:} $p_{AB}$ is the probability of finding $A$ and $B$ simultaneously in the reference area $S_{AB}$ (when a subscript is 0, it represents the absence of the corresponding individual from the reference area); $S_A$ and $S_B$ are the individual areas of A and B, respectively, however they are defined; $n_A$ and $n_B$ are the number of observed fixes of A and B (respectively) in the reference area; $T$ is the number of fixes of each individual. 
	\end{tablenotes}
	\label{pLixn}
\end{table} 

\section{How to define the ellipse of the potential path area}
\label{appendix::Ellipse}
At each step, and for each individual, an ellipse is drawn; two consecutive records are used as focal points and the sum of distances between the focal points and any point of the ellipse is computed as $D = \phi \times \Delta t$ ($\phi$ is the maximum velocity of the individual and $\Delta t$ is the time difference between the consecutive records). The area of the ellipse represents the potential path area \cite{Long2012}.

\section{Metrics derived for each case scenario}
\label{appendix:IndicesTable}

\begin{landscape}
\begin{table}[ht]
\caption{First subset of metrics}
	\centering
    \begin{tabular}{>{\centering\arraybackslash}p{0.3cm}p{0.9cm}p{.9cm}p{.9cm}p{.9cm}p{.8cm}p{.8cm}p{.8cm}p{1cm}p{.8cm}p{1.1cm}p{1.1cm}p{1.1cm}p{.9cm}p{.9cm}p{1cm}p{.8cm}p{.8cm}p{.9cm}p{.9cm}}
		\hline
		& $Prox_{1}$ & $Prox_{2}$ & $Prox_{3}$ & Cs & $HAI_{1}$ & $HAI_{2}$ & $HAI_{3}$ & $L_{ixn}T$ & jPPA & $CSEM_{1}$ & $CSEM_{2}$ & $CSEM_{3}$ & $r_{Lon}$ & $r_{Lat}$ & $r_{Lonlat}$ & $r_{Speed}$ & $DI_{d}$ & $DI_\theta$ & DI \\ 
  \hline
1 & 1.00 & 1.00 & 1.00 & 0.77 & 0.89 & 0.89 & 0.89 & 0.92 & 0.93 & 1.00 & 1.00 & 1.00 & 1.00 & 1.00 & 1.00 & 1.00 & 1.00 & 1.00 & 1.00 \\ 
  2 & 0.70 & 1.00 & 1.00 & 0.61 & 1.00 & 1.00 & 1.00 & 1.00 & 0.86 & 0.56 & 1.00 & 1.00 & 1.00 & 1.00 & 1.00 & 1.00 & 0.82 & 1.00 & 0.82 \\ 
  3 & 0.40 & 0.80 & 1.00 & 0.44 & 1.00 & 1.00 & 1.00 & 1.00 & 0.87 & 0.22 & 0.44 & 1.00 & -1.00 & 1.00 & -0.00 & 1.00 & 1.00 & 0.57 & 0.57 \\ 
  4 & 0.60 & 0.80 & 1.00 & 0.45 & 0.57 & 0.73 & 0.73 & 0.71 & 0.88 & 0.33 & 0.67 & 1.00 & -0.15 & 0.90 & 0.38 & 0.03 & 0.73 & 0.44 & 0.36 \\ 
  5 & 0.60 & 0.80 & 1.00 & -0.15 & 1.00 & 1.00 & 1.00 & 1.00 & 0.90 & 0.44 & 0.56 & 1.00 & -1.00 & -1.00 & -1.00 & 1.00 & 1.00 & -1.00 & -1.00 \\ 
  6 & 0.60 & 0.90 & 1.00 & -0.12 & 0.67 & 0.67 & 0.67 & 0.81 & 0.80 & 0.44 & 0.56 & 1.00 & -0.93 & -0.99 & -0.96 & 1.00 & 0.59 & -1.00 & -0.59 \\ 
  7 & 0.00 & 0.00 & 1.00 & 0.21 &  &  &  &  & 0.67 & 0.00 & 0.00 & 1.00 & 1.00 & 1.00 & 1.00 & 1.00 & 1.00 & 1.00 & 1.00 \\ 
  8 & 0.00 & 0.00 & 0.80 & 0.18 &  &  &  &  & 0.59 & 0.00 & 0.00 & 0.56 & 1.00 & 1.00 & 1.00 & 1.00 & 0.82 & 1.00 & 0.82 \\ 
  9 & 0.60 & 0.60 & 0.80 & 0.35 & 1.00 & 1.00 & 1.00 & 1.00 & 0.68 & 0.56 & 0.56 & 0.78 & -1.00 & 1.00 & 0.00 & 1.00 & 1.00 & 0.79 & 0.79 \\ 
  10 & 0.50 & 0.50 & 0.80 & 0.34 & 0.91 & 0.91 & 0.91 & 0.92 & 0.62 & 0.44 & 0.44 & 0.78 & -0.77 & 0.98 & 0.10 & 0.39 & 0.78 & 0.71 & 0.55 \\ 
  11 & 0.40 & 0.60 & 0.60 & -0.05 & 0.89 & 0.92 & 0.92 & 0.92 & 0.60 & 0.22 & 0.56 & 0.56 & -1.00 & -1.00 & -1.00 & 1.00 & 1.00 & -1.00 & -1.00 \\ 
  12 & 0.50 & 0.60 & 0.60 & -0.03 & 1.00 & 1.00 & 1.00 & 1.00 & 0.64 & 0.44 & 0.56 & 0.56 & -1.00 & -1.00 & -1.00 & 1.00 & 0.82 & -1.00 & -0.82 \\ 
  13 & 0.00 & 0.00 & 0.00 & 0.01 &  &  &  &  & 0.26 & 0.00 & 0.00 & 0.00 & 1.00 & 1.00 & 1.00 & 1.00 & 1.00 & 1.00 & 1.00 \\ 
  14 & 0.00 & 0.00 & 0.00 & 0.09 &  &  &  &  & 0.42 & 0.00 & 0.00 & 0.00 & 1.00 & 1.00 & 1.00 & 1.00 & 0.82 & 1.00 & 0.82 \\ 
  15 & 0.00 & 0.00 & 0.00 & 0.05 &  &  &  &  & 0.29 & 0.00 & 0.00 & 0.00 & -1.00 & 1.00 & 0.00 & 1.00 & 1.00 & 0.79 & 0.79 \\ 
  16 & 0.00 & 0.00 & 0.20 & 0.12 &  &  &  &  & 0.45 & 0.00 & 0.00 & 0.00 & -0.15 & 0.90 & 0.38 & 0.03 & 0.73 & 0.44 & 0.36 \\ 
  17 & 0.00 & 0.20 & 0.60 & -0.03 &  &  &  &  & 0.41 & 0.00 & 0.00 & 0.56 & -1.00 & -1.00 & -1.00 & 1.00 & 1.00 & -1.00 & -1.00 \\ 
  18 & 0.00 & 0.10 & 0.60 & -0.03 &  &  &  &  & 0.44 & 0.00 & 0.00 & 0.56 & -1.00 & -1.00 & -1.00 & 1.00 & 0.82 & -1.00 & -0.82 \\ 
   \hline
	\end{tabular}
\end{table}

\end{landscape}

\section{Summary figures for proximity-speed and proximity-coordination scenarios}
\label{appendix:IndicesFigs}

\setcounter{figure}{0}
\renewcommand{\thefigure}{\ref{appendix:IndicesFigs}.\arabic{figure}}

\begin{figure}[ht!]
	\centering
	\includegraphics[scale=0.45]{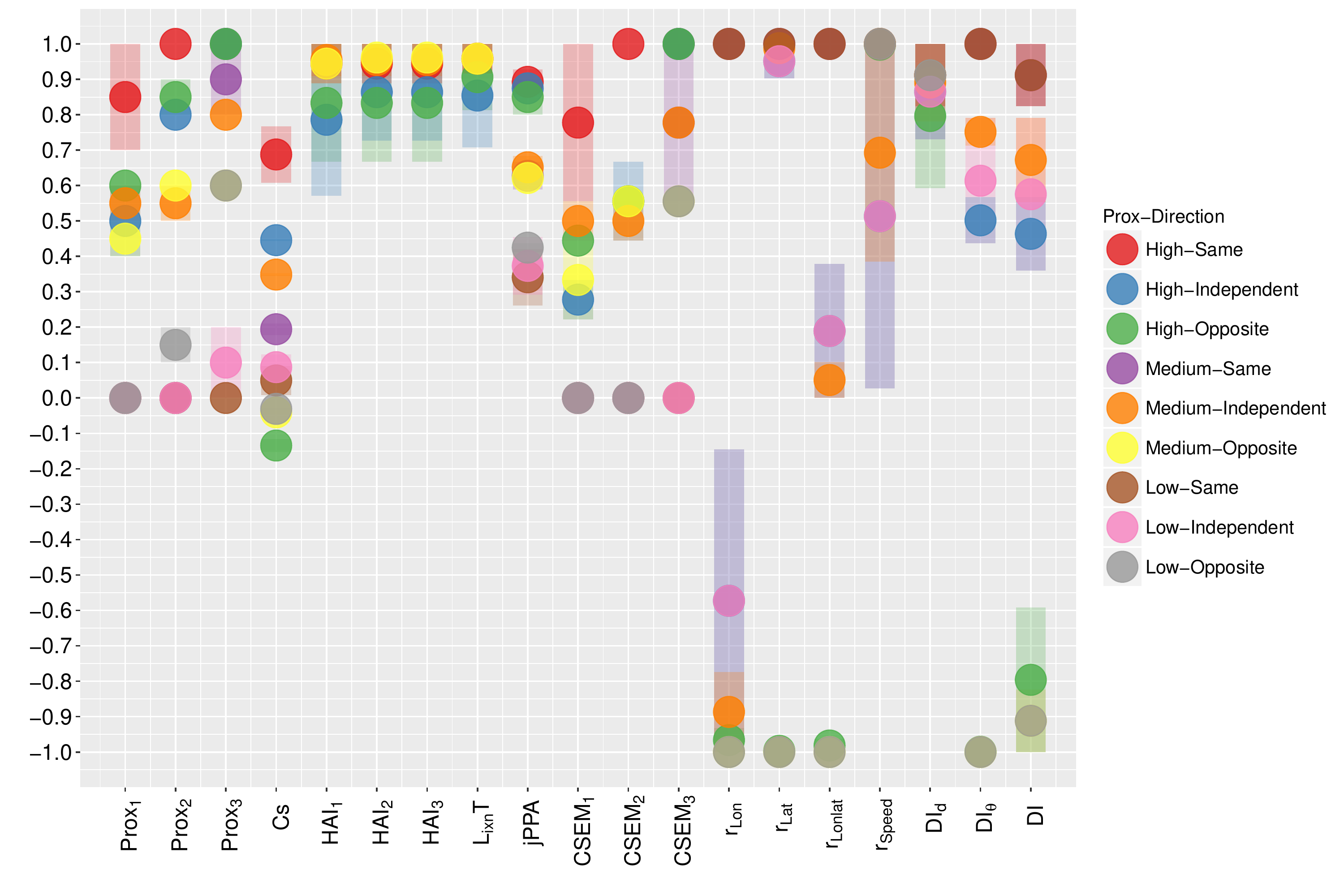}
	\caption{Median (solid circle), minimum and maximum (bar) for each metric by combined category of proximity and direction coordination.}
	\label{ProxDirSensit}
\end{figure}

\begin{figure}[ht!]
	\centering
	\includegraphics[scale=0.45]{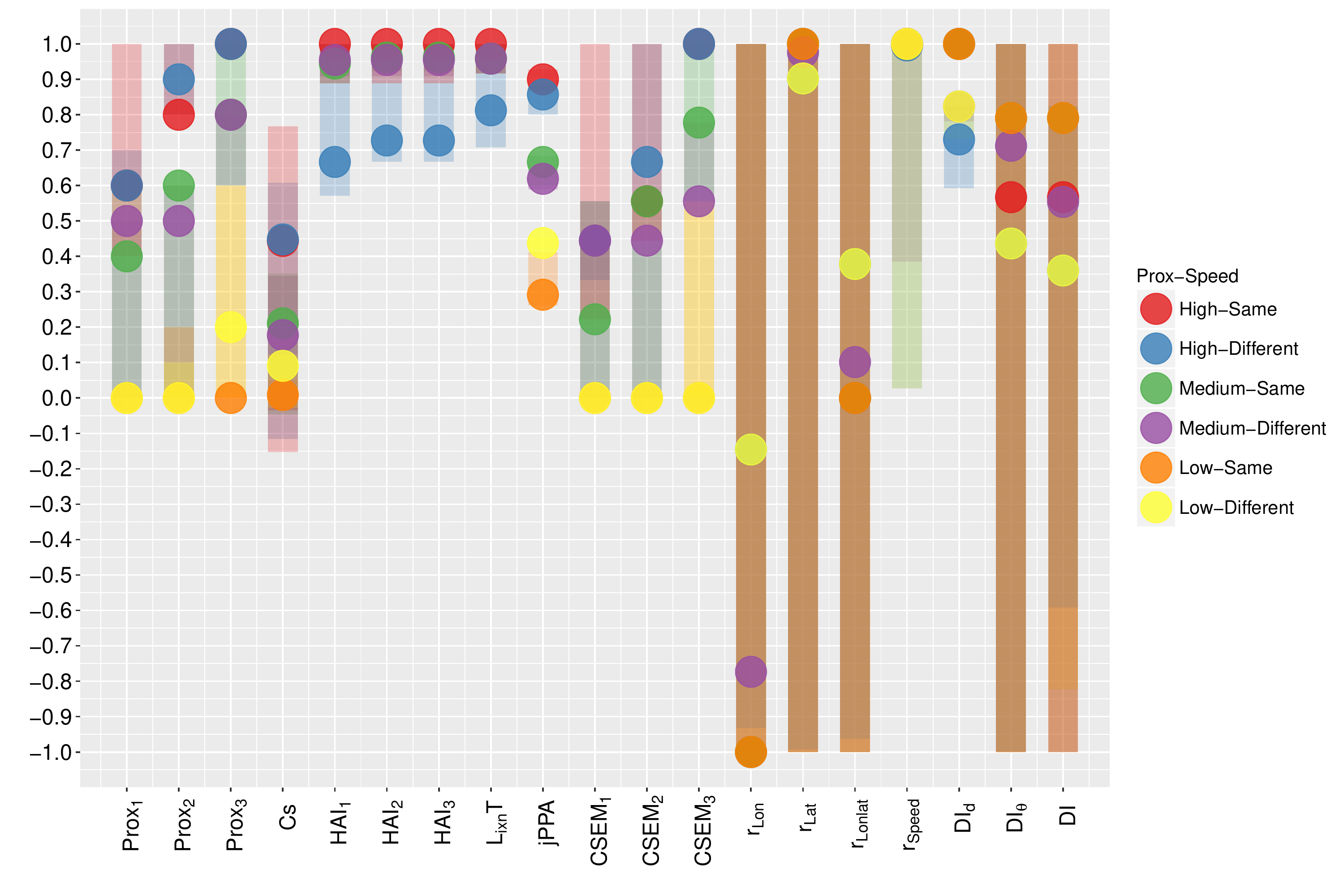}
	\caption{Median (solid circle), minimum and maximum (bar) for each metric by combined category of proximity and speed coordination.}
	\label{ProxSpeedSensit}
\end{figure}

\section{Principal component analysis of the metrics for the case scenarios}
\label{appendix:PCAresults}

\setcounter{figure}{0}
\renewcommand{\thefigure}{\ref{appendix:PCAresults}.\arabic{figure}}

\setcounter{table}{0}
\renewcommand{\thetable}{\ref{appendix:PCAresults}.\arabic{table}}

The initial data table for the PCA is the one in appendix \ref{appendix:IndicesTable}. We discarded HAI and $L_{ixn}T$ because of the missing values. We keep only one of the Prox and CSEM metrics ($Prox_3$ and $CSEM_3$), and kept $r_{Lonlat}$ but not $r_{Lat}$ nor $r_{Lon}$ for the PCA. The final data table was thus composed of 9 variables and 18 individuals. 

\begin{figure}[ht!]
	\centering
	\includegraphics[scale=0.45]{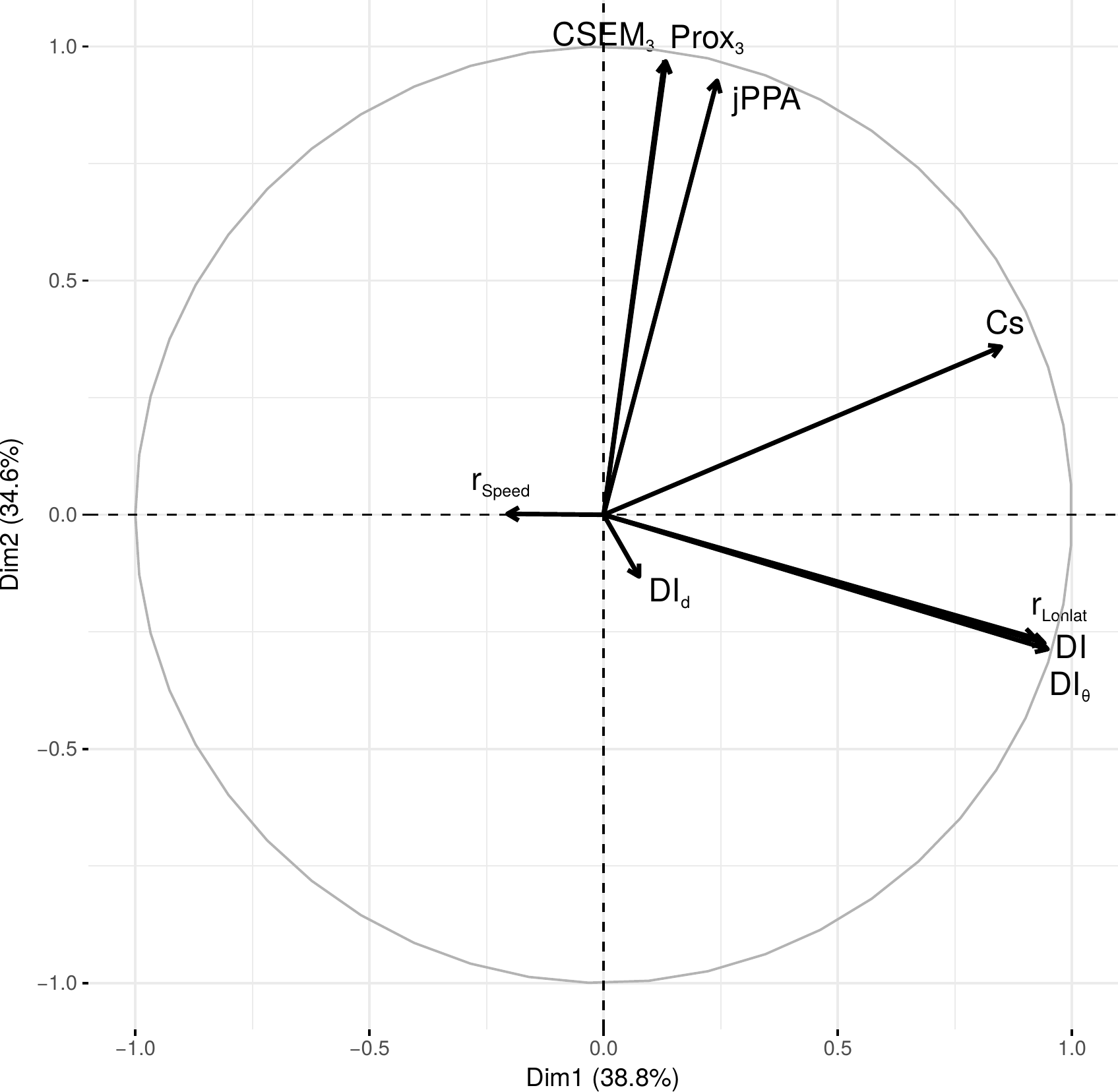}
    	\includegraphics[scale=0.45]{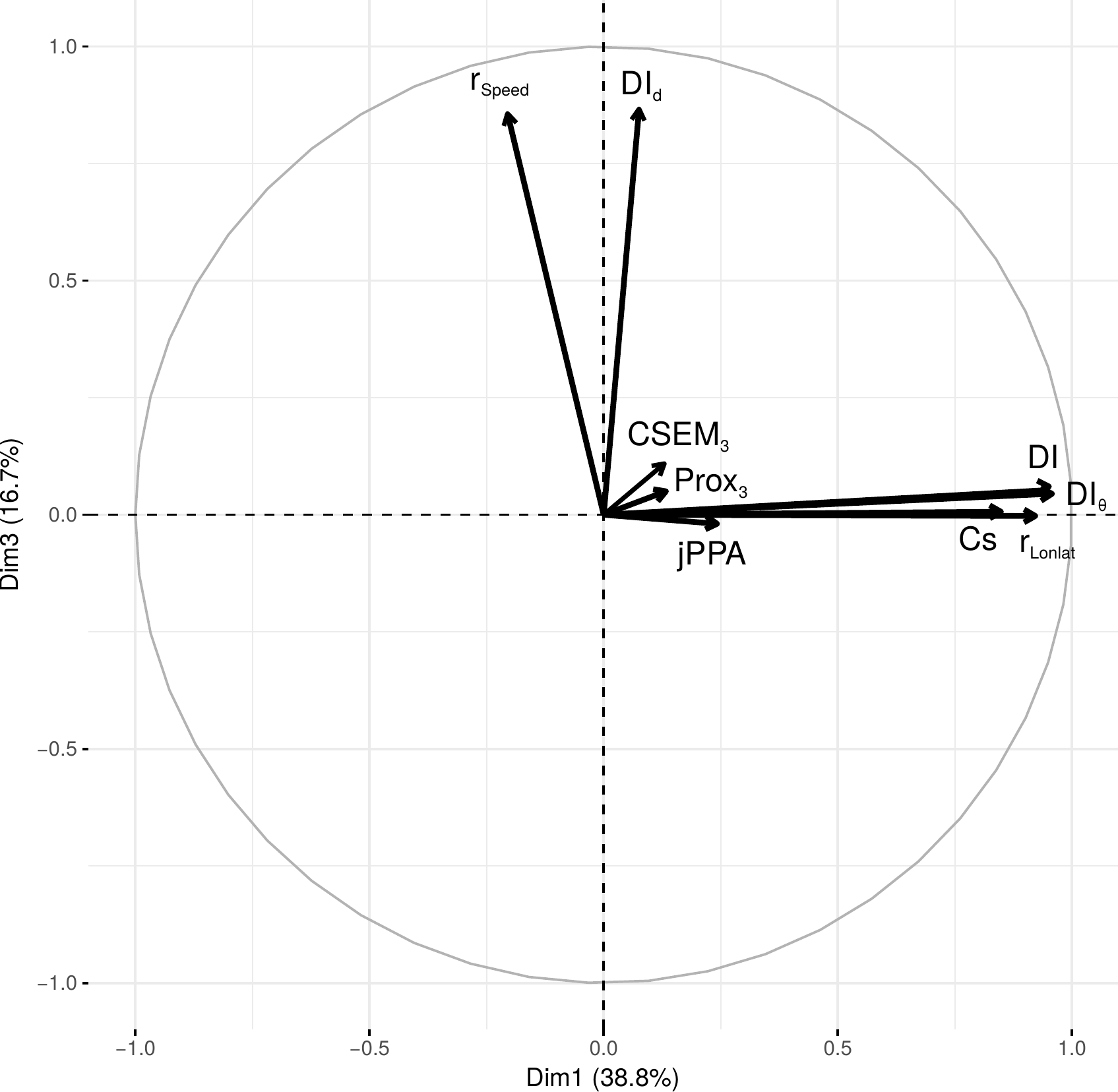}
	\caption{Metric projection in the principal component planes.}
	\label{PCAplot}
\end{figure}

PCA was performed using the FactoMineR package \cite{Husson2013}. We retained 3 components since they explained $90.1\%$ of the total variance. The loadings of each metric regarding each component are detailed in Table \ref{PCAloadings} and represented in Figure \ref{PCAplot}. The first component ($38.8\%$ of the variance) was highly correlated to metrics associated to coordination in direction. The second component ($34.6\%$ of the variance) was highly correlated to proximity-related metrics. The third component ($16.7\%$ of the variance) was highly correlated to the metrics associated to coordination in speed. 

\begin{table}[ht]
\caption{Metric loadings for the three principal components}
\centering
\begin{tabular}{lrrr}
  \hline
 & PC1 & PC2 & PC3 \\ 
  \hline
$Prox_3$ & 0.13 & 0.97 & 0.05 \\ 
  Cs & 0.85 & 0.36 & 0.01 \\ 
  jPPA & 0.24 & 0.93 & -0.02 \\ 
  $CSEM_3$ & 0.13 & 0.97 & 0.11 \\ 
  $r_{Lonlat}$ & 0.92 & -0.26 & -0.00 \\ 
  $r_{Speed}$ & -0.21 & 0.00 & 0.86 \\ 
  $DI_d$ & 0.08 & -0.13 & 0.87 \\ 
  $DI_{\theta}$ & 0.95 & -0.29 & 0.04 \\ 
  DI & 0.94 & -0.27 & 0.05 \\ 
   \hline
\end{tabular}
\label{PCAloadings}
\end{table}

\section{Codes for computing the metrics}
\label{appendix:Codes}

All analyses were performed in R \cite{R2015}. 
Distances between fixes were computed using the pdist package \cite{pdist2013R}. For jPPA calculations, the ellipses were computed as in \cite{Long2014R} and intersection and union areas were approximated by gridding the space via packages polyclip \cite{polyclip2015R} and geoR \cite{geoR2015R}. For $L_{ixn}T$ and HAI, SDMTools \cite{SDMTools2014R} was used to identify points in and out of the reference area. The PCA in appendix \ref{appendix:PCAresults} were performed with the FactoMineR package \cite{FactoMineR}.

Codes with an example, and the dyad tracks arbitrarily created for the case scenarios are accessible from: https://github.com/rociojoo/MetricsDyadJM/

\end{document}